\documentclass[letterpaper]{article} 
\usepackage{aaai2026}  
\usepackage{times}  
\usepackage{helvet}  
\usepackage{courier}  
\usepackage[hyphens]{url}  
\usepackage{graphicx} 
\usepackage{natbib}  
\usepackage{caption} 
\usepackage{cite}
\usepackage{amsmath,amssymb,amsfonts}
\usepackage{textcomp}
\usepackage{multirow}
\usepackage{subcaption}
\usepackage{enumitem}
\usepackage{url}
\usepackage{amsthm}

\newtheorem{definition}{Definition} 
\usepackage{float}
\usepackage{tcolorbox}
\tcbuselibrary{skins} 
\setlist[itemize]{leftmargin=*}
\usepackage{newfloat}
\usepackage{listings}
\DeclareCaptionStyle{ruled}{labelfont=normalfont,labelsep=colon,strut=off} 
\floatstyle{ruled}
\newfloat{listing}{tb}{lst}{}
\floatname{listing}{Listing}
\nocopyright
\title{P-MIA: A Profiled-Based Membership Inference Attack on Cognitive Diagnosis Models}
\author{
    Mingliang Hou\textsuperscript{\rm 1,3}
    Yinuo Wang\textsuperscript{\rm 1},
    Teng Guo\textsuperscript{\rm 1},
    Zitao Liu\textsuperscript{\rm 1}\thanks{The corresponding author.},
    Wenzhou Dou\textsuperscript{\rm 1,3},
    Jiaqi Zheng\textsuperscript{\rm 1},
    Renqiang Luo\textsuperscript{\rm 2},
    Mi Tian\textsuperscript{\rm 3},
    Weiqi Luo\textsuperscript{\rm 1}
}
\affiliations{
    \textsuperscript{\rm 1} Guangdong Institute of Smart Education, Jinan University, Guangzhou, China\\
    \textsuperscript{\rm 2} College of Computer Science and Technology,Jilin University, Changchun, China\\
    \textsuperscript{\rm 3} TAL Education Group, Beijing, China\\

}

\usepackage{bibentry}

\begin{document}

\maketitle

\begin{abstract}
Cognitive diagnosis models (CDMs) are pivotal for creating fine-grained learner profiles in modern intelligent education platforms. However, these models are trained on sensitive student data, raising significant privacy concerns. While membership inference attacks (MIA) have been studied in various domains, their application to CDMs remains a critical research gap, leaving their privacy risks unquantified. This paper is the first to systematically investigate MIA against CDMs. We introduce a novel and realistic grey-box threat model that exploits the explainability features of these platforms, where a model's internal knowledge state vectors are exposed to users through visualizations such as radar charts. We demonstrate that these vectors can be accurately reverse-engineered from such visualizations, creating a potent attack surface. Based on this threat model, we propose a profile-based MIA (P-MIA) framework that leverages both the model's final prediction probabilities and the exposed internal knowledge state vectors as features. Extensive experiments on three real-world datasets against mainstream CDMs show that our grey-box attack significantly outperforms standard black-box baselines. Furthermore, we showcase the utility of P-MIA as an auditing tool by successfully evaluating the efficacy of machine unlearning techniques and revealing their limitations.
\end{abstract}


\section{Introduction}
As artificial intelligence (AI) rapidly advances, the field of education is undergoing a profound paradigm shift from a one-size-fits-all teaching model to personalized learning tailored to individual needs. At the heart of this transformation, cognitive diagnosis models (CDMs) have emerged as a pivotal technology~\cite{survey_cd}. Unlike traditional assessment methods that focus on overall scores, CDMs can delve into students' historical interaction records to provide a fine-grained diagnosis of their mastery level on specific knowledge components (KCs)\footnote{A KC is a generalization of everyday terms like concept, principle, fact, or skill.}~\cite{ecd,neuralcd,graphcd_rcd}. By constructing precise learner profiles, CDMs pave the way for true adaptive learning, enabling dynamic learning path recommendations, timely instructional interventions, and customized practice and feedback for each student, thereby significantly enhancing learning efficiency and experience. Consequently, CDMs have become an indispensable technology in modern online education platforms, intelligent tutoring systems, and large-scale educational assessments.

However, the powerful diagnostic capabilities of CDMs are predicated on vast amounts of highly sensitive personal learning data. Each training instance, typically a triplet of \texttt{\{student ID, question ID, response\}}, directly reflects a specific student's performance on a particular KC~\cite{survey_cd,graphcd_dmccdm}. During the model training process, especially for deep learning models with a large number of parameters, the model may inadvertently ``memorize'' specific patterns from the training data rather than merely learning generalizable rules. This phenomenon gives rise to a critical privacy concern: can a trained CDM leak sensitive information about its training set composition? More specifically, can an adversary, who possesses some of a student's records, determine whether these records were used to train the target model? This problem is known in the security community as a membership inference attack (MIA)~\cite{survey_mia}. This threat is not merely theoretical; it is driven by several real-world motivations. For instance, regulatory bodies might conduct privacy audits via MIA to verify a platform's compliance; a business competitor could perform data-source tracing to ascertain if a rival has used a proprietary or exclusive training dataset; and individual users might seek evidence to support privacy complaints regarding the unauthorized use of their data~\cite{mia_bwbox}.

While MIA have been extensively studied in general machine learning domains such as image classification and natural language processing, their application to the specialized field of CD remains largely unexplored~\cite{mia_bc1,mia_overfitting,mia_xgboost1}. To the best of our knowledge, there is a significant lack of published studies systematically analyzing MIAs against modern CDMs, particularly those based on deep learning. We posit that this research gap stems, in part, from the unique challenges posed by CDMs. Firstly, unlike traditional models that process independent and identically distributed (i.e.d.) samples, CDMs operate on inherently relational interaction triplets, which collectively form a student's learning trajectory. How to effectively define and execute an MIA on such complex, relational data is an open problem in itself. Secondly, due to this data specificity, it is uncertain whether conventional black-box attack methods designed for domains like computer vision can be directly transferred to effectively target CDMs, and what their performance would be. These challenges make privacy risk assessment for CDMs particularly difficult and underscore the urgent need for foundational research in this direction.

To address the aforementioned challenges, we take a different path by focusing on a prevalent trend in modern AI product design: to enhance explainability and user trust, many online education platforms provide visual ``learner profiles''. These seemingly innocuous visualizations are often driven by an internal knowledge state vector (\texttt{kstate\_emb}) generated by the model~\cite{kscd,kancd,neuralcd,neuralncd}. This creates an overlooked grey-box attack surface~\cite{greybox}. Crucially, we find that even if a platform only displays this profile as an image (e.g., a radar chart), an adversary could potentially use advanced computer vision models~\cite{canny} or large language models (LLM)~\cite{llm} to accurately reverse-engineer the underlying numerical knowledge states from the visualization. In this paper, we are the first to model and validate this realistic threat. We propose a novel grey-box MIA framework, the core idea of which is to leverage this exposed knowledge state vector---whether obtained through a direct API call or estimated from an image---as a key feature for the attack model. We argue that the \texttt{kstate\_emb}, as an intermediate representation closer to the model's internal decision logic, exposes the ``memorization'' of training data more significantly than the single, smoothed final prediction probability. Consequently, it can be used to launch attacks that are far more powerful and precise than traditional black-box methods.

The main contributions of this paper can be summarized as follows:
\begin{itemize}
    \item We are the first to systematically investigate MIAs against CDMs, proposing a realistic grey-box threat model that exploits the learner profiles exposed for explainability, even when only visual charts are available.

    \item We design P-MIA, a powerful grey-box attack that significantly outperforms black-box baselines. We then demonstrate its utility as an auditing tool to quantify the limitations of state-of-the-art machine unlearning techniques on CDMs.

    \item We validate our framework across multiple datasets and mainstream CDMs, quantifying the severe privacy risks and providing an in-depth analysis of the underlying causes of the models' vulnerability.
\end{itemize}
\section{Related Work}

\subsection{Cognitive Diagnosis Models}

CDMs are designed to construct a multi-dimensional, explainable cognitive profile for a learner, inferring their mastery of specific KCs from response data~\cite{survey_cd}. Traditional CDMs, rooted in psychometrics like item response theory and the DINA model, rely on manually designed interaction functions with strong theoretical assumptions (e.g., the logistic function)~\cite{irt1,irt2}. While highly interpretable, these fixed functions often fail to capture the complex, non-linear relationships in real-world educational data, thus limiting their diagnostic accuracy. To overcome these limitations, a recent paradigm shift has moved towards deep learning-based CDMs, spearheaded by the NeuralCD framework~\cite{neuralcd,ecd,kancd,kscd,neuralncd}. These models replace handcrafted functions with a multi-layer perceptron (MLP) by projecting students and exercises into embedding spaces to automatically learn their complex interactions. Crucially, to preserve cognitive validity, this framework introduces a monotonicity assumption~\cite{irt1}, ensuring that a student's performance positively correlates with their increasing knowledge mastery. This data-driven paradigm has significantly improved both accuracy and flexibility, inspiring a series of advanced models like KSCD and KaNCD that further explore knowledge associations~\cite{kscd,kancd}.

The core output of these modern CDMs is an explainable knowledge state vector (\texttt{kstate\_emb}), where each dimension directly corresponds to a KC's mastery level~\cite{neuralcd}. In practice, online education platforms translate this vector into user-friendly visualizations, such as competency radar charts or learner profiles, to display a student's cognitive state. This widespread practice of exposing a model's internal state for explainability, however, creates an overlooked information channel. It is precisely this seemingly innocuous learner profile feature that forms the basis of the novel grey-box attack we investigate in this paper.

\subsection{Membership Inference Attacks and Machine Unlearning}

A MIA is a type of privacy attack that aims to determine whether a specific data record was used to train a target model~\cite{survey_mia}. The success of such attacks hinges on a core observation: due to overfitting, machine learning models, especially deep neural networks, exhibit different behaviors on data from their training set (members) versus unseen data (non-members)~\cite{mia_overfitting}. For instance, a model typically yields predictions with higher confidence and lower loss for member data. Attackers exploit this behavioral discrepancy by analyzing a model's output to infer the membership status of a given sample~\cite{mia_icl,mia_bwbox,mia_bc1}. The classic black-box attack paradigm employs ``shadow models'': the attacker trains multiple models that mimic the target's architecture on datasets drawn from a similar distribution, creating a meta-dataset of outputs, and then trains a binary classifier (the attack model) to learn the patterns that distinguish members from non-members~\cite{mia_bc1}.

In response to the privacy risks highlighted by MIAs and regulations like the EU's General Data Protection Regulation (GDPR) with its ``right to be forgotten''~\cite{gdpr}, the field of machine unlearning (MU) has emerged to address the challenge of efficiently and reliably removing the influence of specific data from a trained model~\cite{survey_mu1,survey_mu2}. The most straightforward unlearning method, retraining the model from scratch on the remaining data, is often computationally prohibitive. Consequently, researchers have developed various efficient unlearning algorithms. These methods are broadly categorized into exact unlearning, which aims to produce a model with a parameter distribution identical to that of a retrained model, often through data partitioning schemes like SISA~\cite{mu_sisa}; and approximate unlearning, which seeks a model that is statistically close to the retrained one. Approximate methods, which are more practical for deep models, include techniques based on influence functions, gradient ascent~\cite{mu_ga1,mu_ga2} (negative gradients), and parameter perturbation~\cite{mu_fisher1,mu_fisher2,mu_hessian1,mu_hessian2,mu_ssd}.

Despite the significant attention MIA and MU have received in general machine learning, their intersection with the CD domain remains a critical research gap. To the best of our knowledge, there has been no systematic development of MIAs tailored to modern CDMs, and consequently, no evaluation of the efficacy of MU techniques for these models. This disconnect creates a crucial dilemma: without a reliable ``spear'' (an MIA) to serve as an auditing tool, one cannot measure the strength of any ``shield'' (an MU or other defense mechanism). In other words, developing truly effective privacy-preserving and data-removal mechanisms for CDMs first requires establishing a benchmark that can accurately quantify their membership information leakage risk. This work aims to fill this gap by designing the first MIA paradigm for CDMs, thereby paving the way for future research into validating and developing robust privacy technologies, including machine unlearning, in the field of CD.

\section{Preliminaries}

This section formally defines the core concepts in this paper, and then establishes the specific threat model, including attacker capabilities and goals, that guides our investigation.

\subsection{Definitions}

\begin{definition}
    \textbf{Cognitive Diagnosis}: CD aims to infer a student's proficiency on a set of KCs, $\mathcal{C}=\{c_1, ..., c_K\}$, from their historical responses. Given students $\mathcal{S}$ and exercises $\mathcal{Q}$, a CD model $f_\theta$ learns from student-exercise interactions to estimate the probability of a correct response, $p(r_{ij}=1 | s_i, q_j; \theta)$. Within the model's parameters $\theta$, the student proficiency vector $\boldsymbol{\theta}_{s_i} \in \mathbb{R}^K$ is of particular importance. This vector constitutes the student's underlying knowledge state, which is the direct source for generating the learner profile (e.g., a radar chart) that users interact with. The expert-annotated Q-matrix, $\mathbf{Q} \in \{0, 1\}^{J \times K}$, defines which KCs are required for each exercise.
\end{definition}

\begin{definition}
\label{MIA}
\textbf{Membership Inference Attack}: An MIA is a type of privacy attack that aims to determine whether a given data sample $d=(x, y)$ was part of a target machine learning model's training dataset, denoted as $D_{train}$. This attack exploits the model's tendency to exhibit distinguishable behaviors on its ``members'' ($d \in D_{train}$) versus ``non-members'' ($d \notin D_{train}$). Specifically, a model often yields predictions for its members with higher confidence scores, lower loss values, or smaller prediction entropy. The attacker's goal is to construct an attack function $\mathcal{A}$ that, for any given sample $d$, aims to maximize the accuracy of the following inference:
\begin{equation}
\mathcal{A}(d, M) \rightarrow
\begin{cases}
1 & \text{if } d \in D_{train} \\
0 & \text{if } d \notin D_{train}
\end{cases}
\label{eq:mia_goal}
\end{equation}
where the outputs 1 and 0 represent the inference of member and non-member status, respectively.
\end{definition}

\begin{definition}
\textbf{Machine Unlearning (MU)}: MU is the process of removing the influence of a specific data subset (the forget set, $\mathcal{D}_f$) from a pre-trained model $M_{orig}$ that was trained on the full dataset $\mathcal{D} = \mathcal{D}_r \cup \mathcal{D}_f$, where $\mathcal{D}_r$ is the retain set. The objective is to efficiently produce an unlearned model, $M_{unlearn}$, such that its behavior is computationally indistinguishable from a model $M_{retrain}$ that is retrained from scratch solely on $\mathcal{D}_r$. This process aims to be significantly more efficient than naive retraining.
\end{definition}
MU is a primary and increasingly mainstream defense mechanism against MIA. By effectively removing a sample's contribution, MU aims to nullify the behavioral discrepancies (e.g., higher prediction confidence on members) that MIAs exploit.

\subsection{Attacker Capabilities and Goals}
In this study, we define a threat model that aligns with the standard information hierarchy in machine learning security~\cite{survey_mia,mia_bwbox,greybox}. We establish a baseline black-box attacker who can query the target model with any $(s, q)$ pair to receive the final prediction probability vector (\texttt{predict\_proba}), but has no knowledge of the model's internal workings. Building upon this, we propose a more powerful grey-box attacker. This attacker possesses all the capabilities of the black-box attacker, but is additionally able to obtain or accurately reverse-engineer the student's internal knowledge state vector (\texttt{kstate\_emb}) from the platform's public-facing learner profile. Access to this intermediate model representation elevates the threat to a grey-box setting, and we consider it obtainable either via direct API calls or by reverse-engineering visual charts (e.g., radar charts).

The goal for both attackers is to perform a MIA on $(s, q, r)$ records, motivated by real-world needs such as privacy auditing. Our work aims to demonstrate that the threat posed by a standard black-box attack is dramatically amplified in our proposed grey-box scenario, where the attacker can exploit the internal state information leaked by learner profiles.
\section{The P-MIA Framework}
\begin{figure}[t]
\centering
\includegraphics[width=0.9\columnwidth]{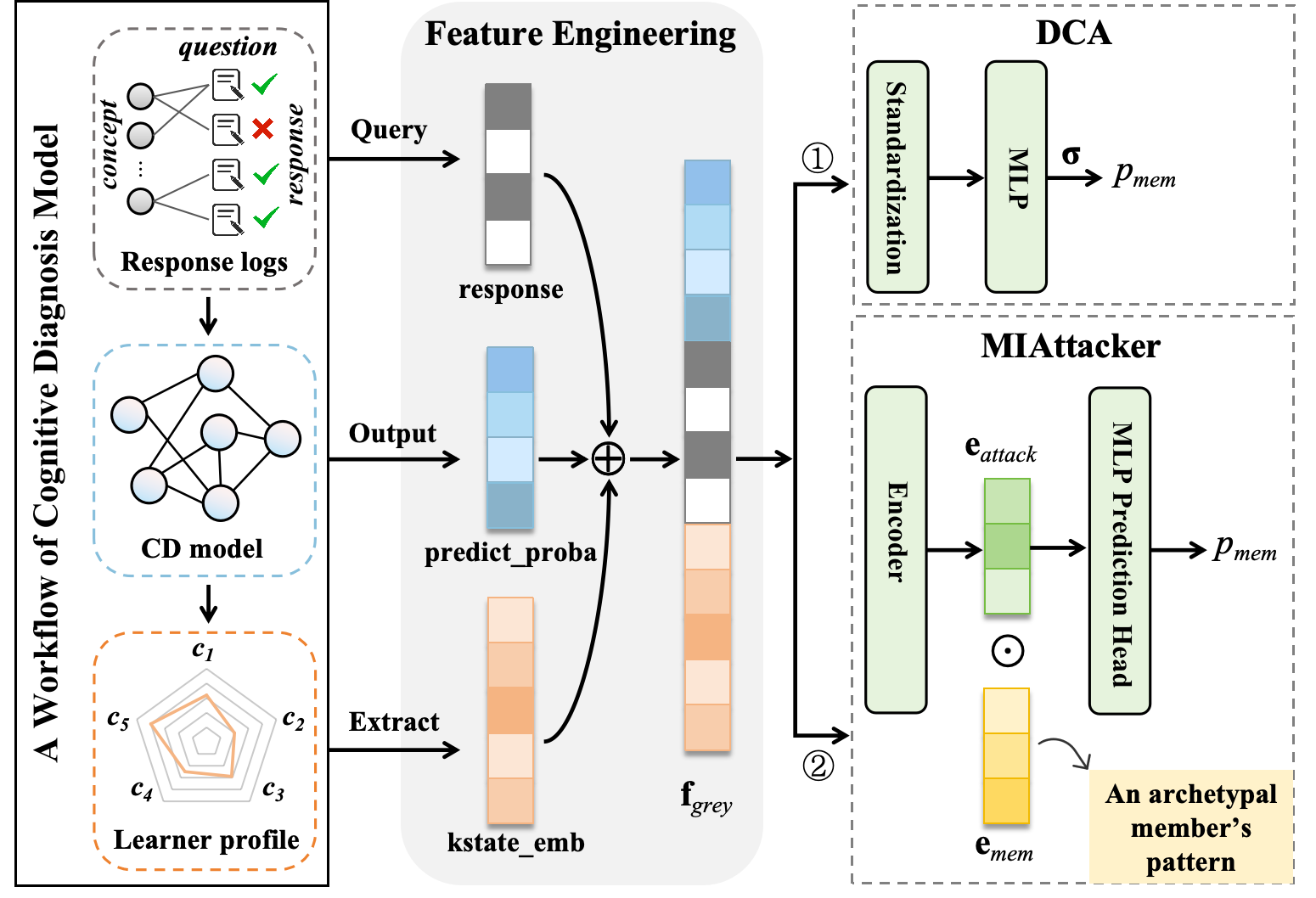} 
\caption{The overall framework of P-MIA.}
\label{framework}
\end{figure}

In this section, we present the technical details of our profile-based MIA (P-MIA). We begin by formally defining the MIA process in the context of CD, followed by a detailed exposition of our feature engineering and attack model architectures (shown in Figure \ref{framework}).
\subsection{Formalizing the MIA Process in CD}
We formulate our MIA as a supervised binary classification task. The attacker's objective is to train an attack model, $\mathcal{A}$, that can distinguish between member and non-member records of a target CD model, $M$. The process involves two conceptual stages: attack model training and attack execution.
\subsubsection{Attack Model Training}The attacker first constructs a training dataset, $\mathcal{D}_{attack}$, to train the attack model $\mathcal{A}$. This dataset is composed of samples for which the attacker has ground-truth membership knowledge. Let $\mathcal{D}_{mem\_train}$ be a set of known member records and $\mathcal{D}_{non-mem\_train}$ be a set of known non-member records.

For each record $d_i \in \mathcal{D}_{mem\_train} \cup \mathcal{D}_{non-mem\_train}$, the attacker queries the target model $M$ to obtain a set of raw signals, $O_i$. Depending on the attacker's capabilities (black-box or grey-box), $O_i$ contains different information. The attacker then applies a feature extraction function, $\Phi$, to transform these signals into a feature vector, $\mathbf{f}_i = \Phi(O_i)$. A membership label, $y_i \in \{1, 0\}$, is assigned to each vector, where $y_i=1$ if $d_i \in \mathcal{D}_{mem\_train}$ and $y_i=0$ otherwise.

The complete attack training dataset is thus $\mathcal{D}_{attack} = \{(\mathbf{f}_i, y_i)\}$. The attacker trains the model $\mathcal{A}$ by minimizing a loss function (e.g., cross-entropy) on this dataset:
\begin{equation}
\min_{\theta_{\mathcal{A}}} \sum_{(\mathbf{f}_i, y_i) \in \mathcal{D}_{attack}} \mathcal{L}(\mathcal{A}(\mathbf{f}_i; \theta_{\mathcal{A}}), y_i)
\label{eq:attack_loss}
\end{equation}
where $\theta_{\mathcal{A}}$ represents the parameters of the attack model.

\subsubsection{Attack Execution}Given a target record $d_{target}$ with unknown membership status, the attacker first queries $M$ to obtain its raw signals $O_{target}$. They then apply the same feature extraction function to get the feature vector $\mathbf{f}_{target} = \Phi(O_{target})$. The trained attack model $\mathcal{A}$ is then used to predict the membership probability:
\begin{equation}
p_{mem} = \mathcal{A}(\mathbf{f}_{target}; \theta_{\mathcal{A}})
\label{eq:attack_predict}
\end{equation}
The attacker infers $d_{target}$ as a member if $p_{mem}$ exceeds a certain threshold (e.g., 0.5).
\subsection{Attack Feature Engineering}
\label{sec:feature_engineering}
The efficacy of our P-MIA is critically dependent on the design of the feature extraction function $\Phi$. We engineer two distinct feature sets for our black-box and grey-box threat models.

The baseline black-box feature vector, $\mathbf{f}_{black}$, is constructed from information readily available from a standard model query: the prediction probability (\texttt{predict\_proba}) and the true response label (\texttt{response}).

Our proposed grey-box attack enhances this baseline by incorporating the student's internal knowledge state vector (\texttt{kstate\_emb}), which we argue is realistically obtainable. We validated that this vector can be reliably reverse-engineered from visual radar charts using a computer vision algorithm based on Canny edge detection to calculate vertex distance ratios~\cite{canny}. This extraction method, along with a tested LLM alternative, proved highly accurate (see Section 5.4 and \textbf{Appendix~B} for details). The final grey-box feature vector is then constructed by concatenating all available information:
\begin{equation}
\mathbf{f}_{grey} = [\texttt{predict\_proba}, \texttt{response}, \texttt{kstate\_emb}]
\label{eq:grey_feature_final}
\end{equation}
We hypothesize this enriched vector is more informative, as the \texttt{kstate\_emb} is a rawer internal representation that more clearly retains the ``fingerprints'' of the training process.
\subsection{Attack Model Architectures}
\label{sec:attack_models}

To validate the effectiveness of our engineered features, we employ two distinct neural network classifiers as our attack model, $\mathcal{A}$. The first is a standard feed-forward classifier, while the second is a custom-designed interactive attacker inspired by CDMs.

\subsubsection{Direct Classification Attacker}We first employ a standard MLP as a strong baseline, which we term the direct classification attacker (DCA). This model learns a direct non-linear mapping from the input feature vector $\mathbf{f}$ (either $\mathbf{f}_{black}$ or $\mathbf{f}_{grey}$) to a membership probability. Prior to training, we standardize the input features. The attack process can be concisely represented as:
\begin{equation}
    p_{mem} = \mathcal{A}_{DCA}(\mathbf{f}; \theta_{DCA}) = \sigma(\text{MLP}(\mathcal{N}(\mathbf{f}); \theta_{DCA}))
\end{equation}
where $\mathcal{N}(\cdot)$ denotes the standardization function applied to the input features, the MLP function consists of two hidden layers (with 64 and 32 neurons, respectively) using ReLU activations, and $\sigma$ is the final sigmoid output function.

\subsubsection{Membership Interaction Attacker}To better capture the nuanced patterns of member samples, we propose the MIAttacker, a novel neural architecture inspired by the interaction mechanisms in CDMs. Instead of directly classifying features, MIAttacker first encodes the input feature vector $\mathbf{f}$ into an attack embedding, $\mathbf{e}_{attack}$. The core of this model is a learnable membership embedding, $\mathbf{e}_{mem}$, which is trained to represent an archetypal member's pattern. An interaction vector, $\mathbf{h}_{int}$, is then produced via an element-wise product:
\begin{equation}
    \mathbf{h}_{int} = \mathbf{e}_{attack} \odot \mathbf{e}_{mem}
\end{equation}
The intuition is that a member sample's embedding will strongly align with $\mathbf{e}_{mem}$, creating a distinctive signal. This interaction vector is then passed to a final MLP prediction head to output the membership probability, $p_{mem}$. The entire model is trained end-to-end.

\section{Experiments}
\label{sec:experiments}
We conduct a series of experiments to comprehensively evaluate our P-MIA framework. Our evaluation is designed to: (1) demonstrate the significant performance gap between our grey-box attack and a standard black-box baseline; (2) quantify the contribution of the \texttt{kstate\_emb} through ablation and sensitivity analyses; and (3) showcase P-MIA's utility as an effective tool for auditing state-of-the-art machine unlearning defenses.
\begin{table}[ht]
\centering
\renewcommand{\arraystretch}{1.2} 
\begin{tabular}{ccccc}
\hline
\textbf{Dataset} & \textbf{\#Student} & \textbf{\#Question} & \textbf{\#Response} & \textbf{\#KC} \\ \hline
Math1            & 4,209              & 20                  & 84,180              & 11            \\
Math2            & 3,911              & 20                  & 78,220              & 16            \\
Frcsub           & 536                & 20                  & 10,720              & 8             \\ \hline
\end{tabular}
\caption{Statistics of the datasets used in our experiments.}
\label{dataset}
\end{table}

\begin{table*}[ht]
\renewcommand{\arraystretch}{1.2}
\centering
\resizebox{\textwidth}{!}{%
\begin{tabular}{cc|cccccc|cccccc|cccccc}
\hline
\multicolumn{2}{c|}{\textbf{Dataset}} &
  \multicolumn{6}{c|}{\textbf{Math1}} &
  \multicolumn{6}{c|}{\textbf{Math2}} &
  \multicolumn{6}{c}{\textbf{Frcsub}} \\ \hline
\multicolumn{2}{c|}{\textbf{Target w/ Defenses}} &
  \multicolumn{2}{c}{\textbf{XGBoost}} &
  \multicolumn{2}{c}{\textbf{DCA}} &
  \multicolumn{2}{c|}{\textbf{MIAttacker}} &
  \multicolumn{2}{c}{\textbf{XGBoost}} &
  \multicolumn{2}{c}{\textbf{DCA}} &
  \multicolumn{2}{c|}{\textbf{MIAttacker}} &
  \multicolumn{2}{c}{\textbf{XGboost}} &
  \multicolumn{2}{c}{\textbf{DCA}} &
  \multicolumn{2}{c}{\textbf{MIAttacker}} \\ \hline
\textbf{Target Model} &
  \textbf{Defenses} &
  ACC &
  AUC &
  ACC &
  AUC &
  ACC & 
  AUC &
  ACC &
  AUC &
  ACC &
  AUC &
  ACC &
  AUC &
  ACC &
  AUC &
  ACC &
  AUC &
  ACC &
  AUC \\ \hline
\multirow{4}{*}{NeuralCD} &
  * &
  0.7605 &
  0.8194 &
  1.0000 &
  1.0000 &
  0.9883 &
  0.9947 &
  0.7474 &
  0.7967 &
  0.9929 &
  0.9947 &
  0.9835 &
  0.9800 &
  0.7085 &
  0.7891 &
  0.9955 &
  1.0000 &
  0.9417 &
  0.9807 \\
 &
  Amnesiac &
  0.7644 &
  0.8333 &
  1.0000 &
  1.0000 &
  0.9907 &
  0.9993 &
  0.7443 &
  0.7877 &
  0.9929 &
  0.9947 &
  0.9898 &
  0.9892 &
  0.6951 &
  0.7575 &
  0.9955 &
  1.0000 &
  1.0000 &
  1.0000 \\
 &
  L-CODEC &
  0.6423 &
  0.7026 &
  0.9922 &
  0.9999 &
  0.9658 &
  0.9844 &
  0.6475 &
  0.6860 &
  0.9929 &
  0.9947 &
  0.9929 &
  0.9999 &
  0.6054 &
  0.6811 &
  0.9417 &
  0.9997 &
  0.9103 &
  0.9755 \\
 &
  SSD &
  \textit{0.6236} &
  \textit{0.6497} &
  \textbf{0.9666} &
  \textbf{0.9959} &
  0.9463 &
  0.9821 &
  \textit{0.5917} &
  \textit{0.5910} &
  \textbf{0.9795} &
  \textbf{0.9939} &
  0.9630 &
  0.9725 &
  \textit{0.4843} &
  \textit{0.4635} &
  0.8386 &
  0.9997 &
  \textbf{0.9327} &
  \textbf{0.9768} \\ \hline
\multirow{4}{*}{KSCD} &
  * &
  0.8538 &
  0.8924 &
  0.9953 &
  1.0000 &
  0.9953 &
  1.0000 &
  0.7608 &
  0.8258 &
  0.9559 &
  0.9825 &
  0.9591 &
  0.9848 &
  0.6502 &
  0.7260 &
  0.9193 &
  0.9481 &
  0.9148 &
  0.9294 \\
 &
  Amnesiac &
  0.8530 &
  0.8945 &
  0.9953 &
  1.0000 &
  0.9930 &
  1.0000 &
  0.7624 &
  0.8219 &
  0.9559 &
  0.9818 &
  0.9599 &
  0.9841 &
  0.6502 &
  0.7260 &
  0.9193 &
  0.9481 &
  0.9238 &
  0.9141 \\
 &
  L-CODEC &
  0.8281 &
  0.8655 &
  0.9642 &
  0.9638 &
  0.9557 &
  0.9584 &
  0.6869 &
  0.7590 &
  0.8985 &
  0.9116 &
  0.9048 &
  0.9335 &
  \textit{0.5022} &
  \textit{0.4547} &
  \textbf{0.5157} &
  \textbf{0.3992} &
  0.4978 &
  0.4411 \\
 &
  SSD &
  \textit{0.7325} &
  \textit{0.7666} &
  \textbf{0.8523} &
  \textbf{0.9111} &
  0.8320 &
  0.9066 &
  \textit{0.5618} &
  \textit{0.5648} &
  \textbf{0.6845} &
  \textbf{0.5808} &
  0.6633 &
  0.6085 &
  0.4529 &
  0.4544 &
  0.5695 &
  0.5105 &
  0.5740 &
  0.4587 \\ \hline
\multirow{4}{*}{KaNCD} &
  * &
  0.7146 &
  0.7826 &
  0.8974 &
  0.9347 &
  0.8911 &
  0.9398 &
  0.7608 &
  0.8258 &
  0.9559 &
  0.9825 &
  0.9591 &
  0.9848 &
  0.6323 &
  0.7268 &
  0.8296 &
  0.9286 &
  0.7668 &
  0.8008 \\
 &
  Amnesiac &
  0.6843 &
  0.7424 &
  0.8958 &
  0.9412 &
  0.8911 &
  0.9437 &
  0.7624 &
  0.8219 &
  0.9559 &
  0.9818 &
  0.9599 &
  0.9841 &
  0.6368 &
  0.6997 &
  0.8341 &
  0.9325 &
  0.7623 &
  0.7958 \\
 &
  L-CODEC &
  0.5630 &
  0.5989 &
  0.8281 &
  0.9590 &
  0.7908 &
  0.8127 &
  0.6869 &
  0.7590 &
  0.8985 &
  0.9116 &
  0.9048 &
  0.9335 &
  \textit{0.4260} &
  \textit{0.5599} &
  0.4843 &
  0.4338 &
  \textbf{0.5112} &
  \textbf{0.4690} \\
 &
  SSD &
  \textit{0.5653} &
  \textit{0.5529} &
  \textbf{0.6330} &
  \textbf{0.5195} &
  0.6283 &
  0.5147 &
  \textit{0.5618} &
  \textit{0.5648} &
  \textbf{0.6845} &
  \textbf{0.5808} &
  0.6633 &
  0.6085 &
  0.4350 &
  0.3961 &
  0.4978 &
  0.4667 &
  0.4888 &
  0.5094 \\ \hline
\end{tabular}%
}
\caption{MIA results on all datasets for the 5\% forgetting ratio scenario. We compare the black-box (XGBoost) and grey-box (DCA, MIAttacker) attacks. `*' denotes no defense. \textit{Italicized} values mark the most effective defense against the black-box attack, while \textbf{bold} values highlight the best performance for each of our grey-box attackers within a model block.}
\label{mainres}
\end{table*}

\begin{table*}[ht]
\centering
\renewcommand{\arraystretch}{1.1}
\resizebox{\textwidth}{!}{%
\begin{tabular}{cc|cccc|cccc|cccc}
\hline
\multicolumn{2}{c|}{\textbf{Dataset}} &
  \multicolumn{4}{c|}{\textbf{Math1}} &
  \multicolumn{4}{c|}{\textbf{Math2}} &
  \multicolumn{4}{c}{\textbf{Frcsub}} \\ \hline
\multicolumn{2}{c|}{\textbf{Target w/ Defenses}} &
  \multicolumn{2}{c}{\textbf{\begin{tabular}[c]{@{}c@{}}DCA w/o \\ kstate\_emb\end{tabular}}} &
  \multicolumn{2}{c|}{\textbf{\begin{tabular}[c]{@{}c@{}}MIAttacker w/o\\  kstate\_emb\end{tabular}}} &
  \multicolumn{2}{c}{\textbf{\begin{tabular}[c]{@{}c@{}}DCA w/o \\ kstate\_emb\end{tabular}}} &
  \multicolumn{2}{c|}{\textbf{\begin{tabular}[c]{@{}c@{}}MIAttacker w/o\\  kstate\_emb\end{tabular}}} &
  \multicolumn{2}{c}{\textbf{\begin{tabular}[c]{@{}c@{}}DCA w/o \\ kstate\_emb\end{tabular}}} &
  \multicolumn{2}{c}{\textbf{\begin{tabular}[c]{@{}c@{}}MIAttacker w/o \\ kstate\_emb\end{tabular}}} \\ \hline
\textbf{Target Model} &
  \textbf{Defenses} &
  $\text{ACC}_{\text{MIA}}$ &
  $\text{AUC}_{\text{MIA}}$ &
  $\text{ACC}_{\text{MIA}}$ &
  $\text{AUC}_{\text{MIA}}$ &
  $\text{ACC}_{\text{MIA}}$ &
  $\text{AUC}_{\text{MIA}}$ &
  $\text{ACC}_{\text{MIA}}$ &
  $\text{AUC}_{\text{MIA}}$ &
  $\text{ACC}_{\text{MIA}}$ &
  $\text{AUC}_{\text{MIA}}$ &
  $\text{ACC}_{\text{MIA}}$ &
  $\text{AUC}_{\text{MIA}}$ \\ \hline
\multirow{4}{*}{NeuralCD} &
  * &
  \textbf{0.6493} &
  \textbf{0.7073} &
  0.5311 &
  0.5386 &
  0.6562 &
  0.6952 &
  0.6239 &
  0.6341 &
  0.6906 &
  0.7261 &
  0.5964 &
  0.7030 \\
 &
  Amnesiac &
  0.6470 &
  0.7067 &
  0.5202 &
  0.5462 &
  0.6640 &
  0.7145 &
  0.6223 &
  0.6408 &
  0.6816 &
  0.7221 &
  0.6278 &
  0.7091 \\
 &
  L-CODEC &
  0.5443 &
  0.5214 &
  0.3398 &
  0.3246 &
  0.7235 &
  \textbf{0.9052} &
  \textbf{0.7494} &
  0.8758 &
  0.6457 &
  \textbf{0.8221} &
  0.7175 &
  \textbf{0.8998} \\
 &
  SSD &
  0.5412 &
  0.5359 &
  0.4977 &
  0.4876 &
  0.4894 &
  0.4809 &
  0.4854 &
  0.4692 &
  0.5785 &
  0.6643 &
  0.7534 &
  0.7625 \\ \hline
\multirow{4}{*}{KSCD} &
  * &
  0.6306 &
  \textbf{0.6768} &
  0.5474 &
  \textbf{0.6335} &
  \textbf{0.6074} &
  \textbf{0.6298} &
  \textbf{0.5885} &
  \textbf{0.5884} &
  \textbf{0.6726} &
  \textbf{0.7477} &
  0.6323 &
  \textbf{0.7128} \\
 &
  Amnesiac &
  \textbf{0.6314} &
  0.6765 &
  \textbf{0.5568} &
  0.6326 &
  0.6066 &
  0.6298 &
  0.5625 &
  0.5837 &
  0.6726 &
  0.7468 &
  \textbf{0.6368} &
  0.7082 \\
 &
  L-CODEC &
  0.5529 &
  0.5446 &
  0.3095 &
  0.3696 &
  0.4485 &
  0.3878 &
  0.3682 &
  0.3426 &
  0.3946 &
  0.2040 &
  0.3408 &
  0.4251 \\
 &
  SSD &
  0.5443 &
  0.5439 &
  0.4868 &
  0.5169 &
  0.5366 &
  0.5339 &
  0.5122 &
  0.5092 &
  0.4126 &
  0.3195 &
  0.4305 &
  0.4759 \\ \hline
\multirow{4}{*}{KaNCD} &
  * &
  \textbf{0.6260} &
  \textbf{0.6526} &
  0.5537 &
  0.5989 &
  \textbf{0.5696} &
  0.5342 &
  \textbf{0.5342} &
  \textbf{0.5314} &
  0.6996 &
  \textbf{0.7493} &
  \textbf{0.6368} &
  \textbf{0.7231} \\
 &
  Amnesiac &
  0.6221 &
  0.6406 &
  \textbf{0.5747} &
  \textbf{0.6170} &
  0.5547 &
  \textbf{0.5856} &
  0.5208 &
  0.5223 &
  \textbf{0.7040} &
  0.7490 &
  0.6099 &
  0.7201 \\
 &
  L-CODEC &
  0.4409 &
  0.3973 &
  0.3149 &
  0.2832 &
  0.3714 &
  0.3548 &
  0.3320 &
  0.3056 &
  0.3812 &
  0.1692 &
  0.2556 &
  0.2750 \\
 &
  SSD &
  0.5420 &
  0.5424 &
  0.5016 &
  0.5140 &
  0.5287 &
  0.5713 &
  0.4957 &
  0.5199 &
  0.3812 &
  0.2535 &
  0.3004 &
  0.3341 \\ \hline
\end{tabular}%
}
\caption{Ablation study results showing the performance of our grey-box attackers (DCA and MIAttacker) when the knowledge state vector is withheld (w/o \texttt{kstate\_emb}). This setup degrades them to black-box attackers. All experiments are conducted with a 5\% forgetting ratio. \textbf{Bold} values indicate the best performance for each model-dataset block under this ablated setting.}
\label{abres}
\end{table*}
\subsection{Experimental Settings}
\subsubsection{Datasets}We conduct our experiments on three real-world, publicly available educational datasets: Math1, Math2, and FrcSub~\cite{survey_cd}. The key statistics of these datasets are summarized in Table~\ref{dataset}. Specifically, Math1 and Math2 originate from large-scale mathematics examinations for high school students, covering a diverse set of KCs in topics such as algebra and geometry. In contrast, FrcSub centers on a more specialized domain, containing the response logs of middle school students on fraction subtraction problems. Each dataset provides the necessary student response records and their corresponding Q-matrices. For further in-depth descriptions of each dataset, please refer to \textbf{Appendix A.1}.
\subsubsection{Baselines}Our experimental baselines are structured around three categories: the target CDMs we attack, the attack models we employ, and the unlearning-based defenses we audit. \textbf{(1) Target CDMs:} We conduct our attacks against three representative deep learning-based CDMs: NeuralCD~\cite{neuralcd}, KSCD~\cite{kscd}, and KaNCD~\cite{kancd}. \textbf{(2) Attack Models:} To establish a performance benchmark, we implement a strong black-box baseline using an XGBoost classifier, which is trained exclusively on the limited feature set of $[\texttt{predict\_proba}, \texttt{response}]$. In contrast, our proposed grey-box attacks are implemented using the DCA and MIAttacker architectures, both of which leverage the enriched feature set $[\texttt{predict\_proba}, \texttt{response}, \texttt{kstate\_emb}]$. \textbf{(3) Defenses for Auditing:} To demonstrate the utility of P-MIA as an auditing tool, we evaluate its effectiveness against models protected by three state-of-the-art classes of approximate MU models: a gradient ascent-based method (Amnesiac)~\cite{mu_ga1}, a Hessian-based method (L-CODEC)~\cite{mu_hessian1}, and a Fisher information matrix (FIM)-based method (SSD)~\cite{mu_ssd}. Detailed descriptions of each baseline method are provided in \textbf{Appendix A.2}.
\subsubsection{Evaluation Metrics}We assess the performance of our P-MIA framework using standard metrics for binary classification, as is common practice in MIA literature~\cite{survey_mia}. Our primary evaluation metrics are \textbf{Attack Accuracy ($\text{ACC}_{\text{MIA}}$)} and \textbf{Attack AUC ($\text{AUC}_{\text{MIA}}$)}, where a score significantly above 0.5 indicates a successful attack.

\begin{figure}[ht]
\centering
\includegraphics[width=1.0\columnwidth]{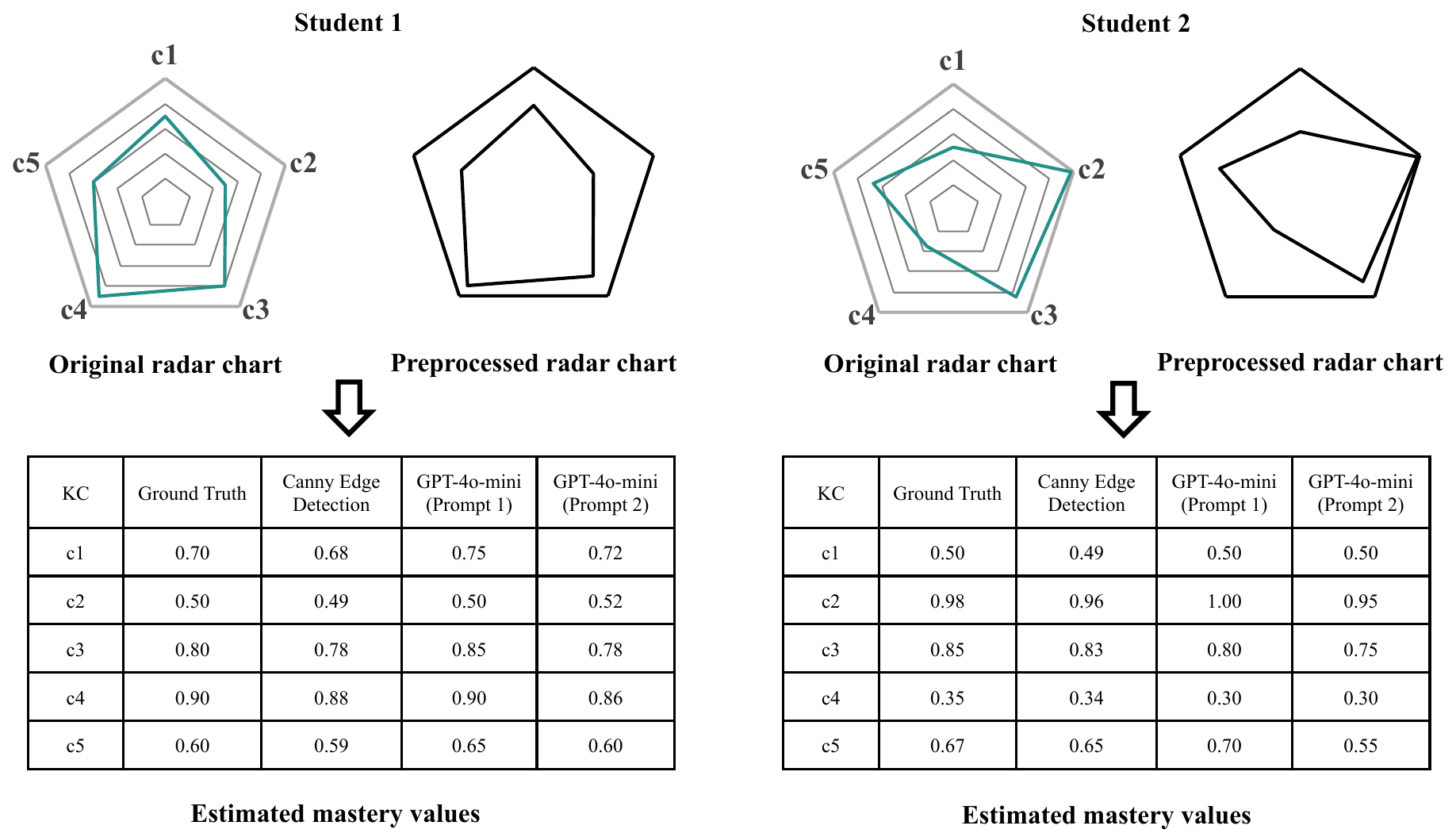} 
\caption{Demonstration of the high-fidelity extraction of a student's \texttt{kstate\_emb} from a visual radar chart. The process involves preprocessing the image to isolate contours (top panel). A quantitative analysis (bottom panel) shows that our Canny Edge Detection method achieves very low error compared to the ground truth. Estimates from an LLM (GPT-4o-mini) are also shown for comparison.}
\label{vs_radar}
\end{figure}
\begin{figure*}[ht]
		\centering
		\begin{minipage}[t]{0.31\textwidth}
			\centering
			\includegraphics[width=5.5cm]{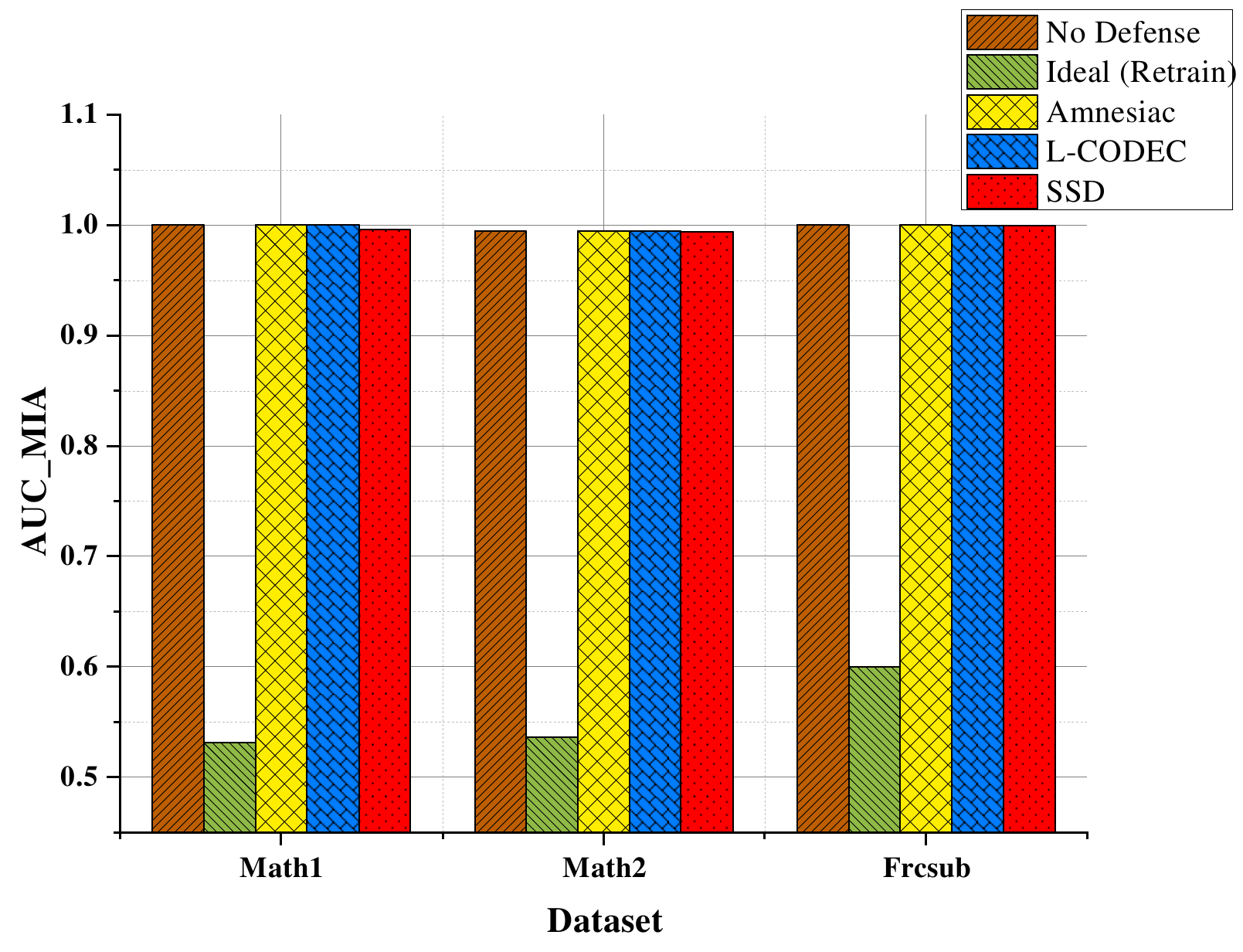}
			\subcaption{NeuralCD}
		\end{minipage}
		\begin{minipage}[t]{0.31\textwidth}
			\centering
			\includegraphics[width=5.5cm]{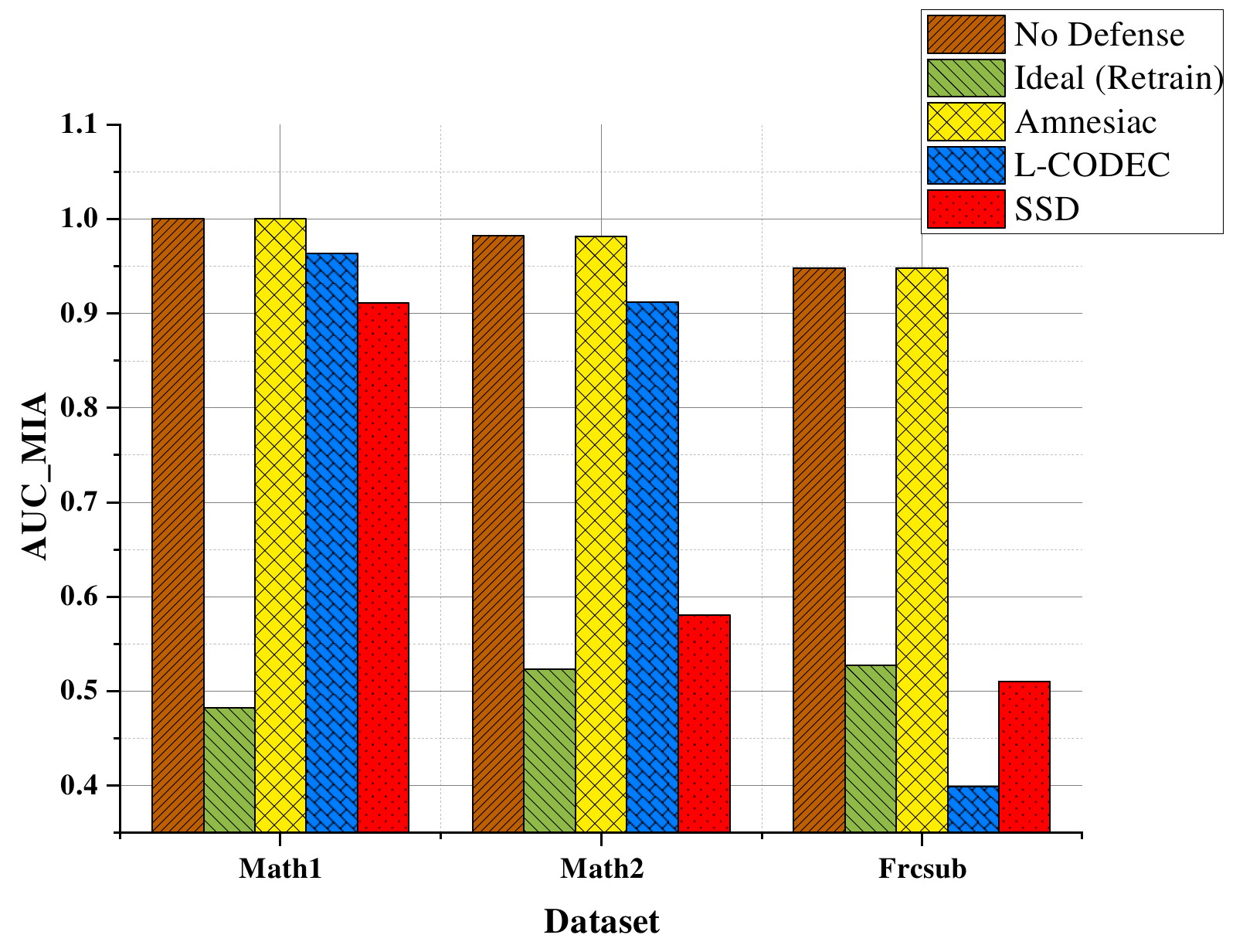}
			\subcaption{KSCD}
		\end{minipage}
		\begin{minipage}[t]{0.31\textwidth}
			\centering
			\includegraphics[width=5.5cm]{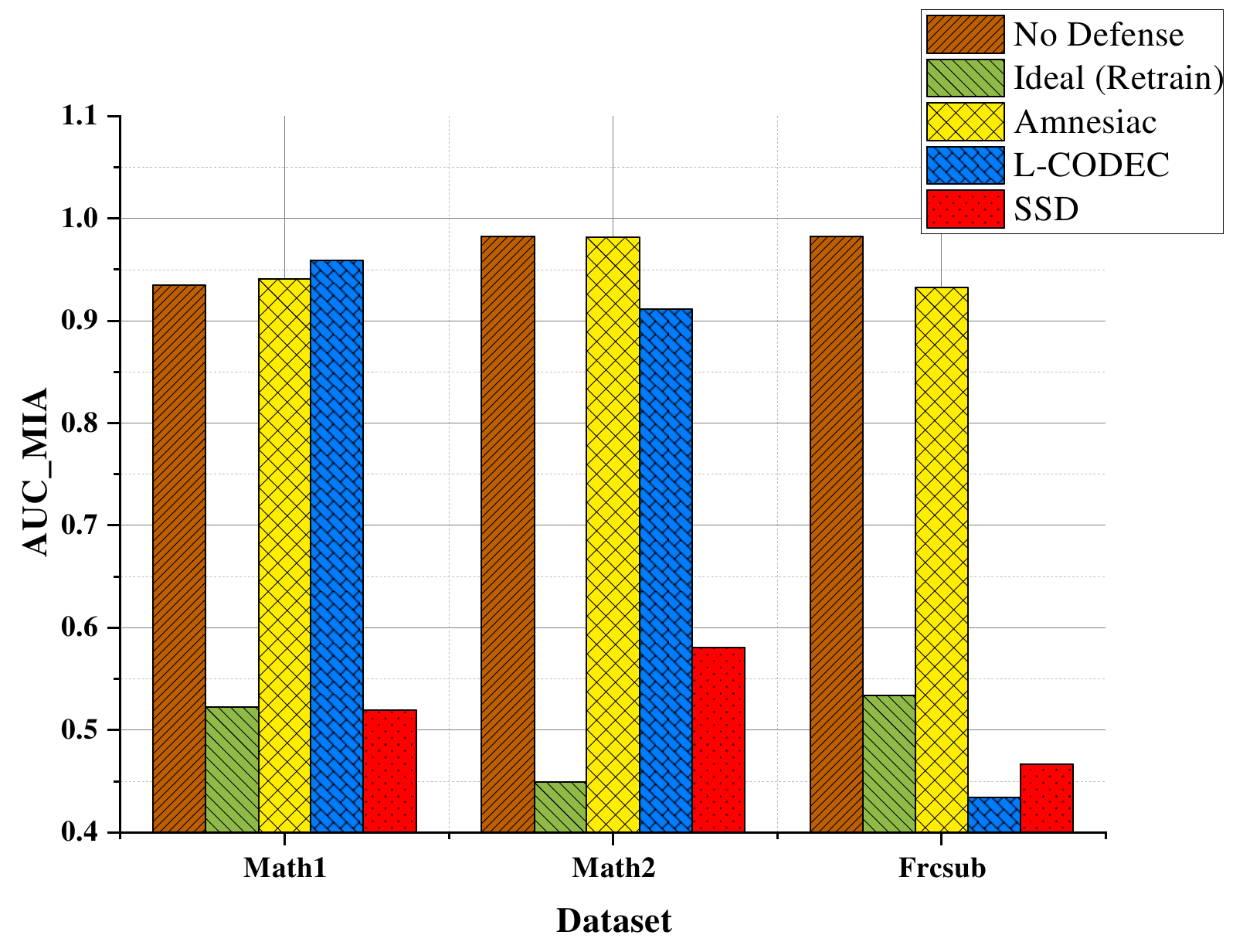}
			\subcaption{KaNCD}
		\end{minipage}
		\caption{Audit of machine unlearning defenses using our P-MIA (MIAttacker) with a 5\% forgetting ratio. Each subplot shows the attack AUC\textsubscript{MIA} against a different target CDM.}
		\label{vs}
\end{figure*}

Our evaluation protocol is designed not only to assess the attack's efficacy but also to audit unlearning defenses. The core methodology involves training the attack model on a dataset of known members and non-members, and then evaluating it on a disjoint set of samples under different defense conditions (e.g., against an original vs. a retrained model). This comprehensive approach allows us to quantify both the direct privacy leakage and the effectiveness of mitigation strategies. The formal definitions of all metrics and a detailed, step-by-step description of this protocol are provided in \textbf{Appendix A.3}.

The full reproducible code for our experiments is publicly available on GitHub\footnote{\url{https://anonymous.4open.science/r/p-mia-6771}}. Further implementation details can be found in \textbf{Appendix A.4}. For the MIA-related experiments, we evaluate scenarios where the data forgetting ratios are set to 10\%, 5\%, and 1\%. Due to space constraints in the main paper, we primarily report the results for the 5\% ratio scenario. The complete results for all ratios are provided in \textbf{Appendix B}.

\subsection{P-MIA Efficacy}
The comprehensive results presented in Table~\ref{mainres} lead to several key findings. 

First and most strikingly, our proposed grey-box attackers (DCA and MIAttacker) consistently and overwhelmingly outperform the black-box (XGBoost) baseline across all datasets and target models. For instance, on the undefended NeuralCD model with the Math1 dataset, the grey-box attackers achieve perfect or near-perfect AUC\textsubscript{MIA} scores ($\approx 1.0000$), whereas the black-box attacker only reaches 0.8194. This massive performance gap empirically validates our core hypothesis: the knowledge state vector (\texttt{kstate\_emb}) serves as a potent information channel that dramatically amplifies membership inference threats.

Second, while machine unlearning defenses show some efficacy, they are largely insufficient against our powerful grey-box P-MIA framework. Against the black-box attacker, defenses like SSD can significantly degrade attack performance, as seen on KaNCD with the Frcsub dataset where the AUC\textsubscript{MIA} drops from 0.7268 to 0.3961. However, these defenses offer minimal protection against grey-box attackers. For example, when defending NeuralCD on Math2 with SSD, the DCA's AUC\textsubscript{MIA} remains exceptionally high at 0.9939, demonstrating that the information leaked through the \texttt{kstate\_emb} allows the attack to bypass the protection offered by these unlearning methods.

Third, our MIAttacker, which is specifically designed to model feature interactions, often demonstrates the most resilient attack performance against the strongest defenses. For example, on the defended KaNCD model with the Math2 dataset, MIAttacker achieves the highest AUC\textsubscript{MIA} of 0.9879 against the L-CODEC defense, outperforming both the DCA (0.0807) and XGBoost (0.5034) in this challenging scenario. This suggests that its specialized architecture is particularly effective at identifying the subtle membership fingerprints that remain even after unlearning. Overall, our results not only quantify the severe privacy risks posed by learner profiles but also establish P-MIA as a robust tool for auditing the practical limitations of current unlearning defenses.

\subsection{Quantifying the Impact of the Knowledge State Vector}
To empirically validate our central hypothesis that the exposed knowledge state vector (\texttt{kstate\_emb}) is the primary driver of the grey-box attack's power, we conduct a critical ablation study. In this study, we evaluate the performance of our two proposed grey-box attackers, DCA and MIAttacker, under a ``knowledge-blind'' condition where the \texttt{kstate\_emb} is deliberately removed from their feature set. This effectively degrades them to black-box attackers that rely solely on the $[\texttt{predict\_proba}, \texttt{response}]$ features. The goal is to isolate and quantify the performance contribution of the learner profile information.

The results of this ablation are presented in Table~\ref{abres} (w/o \texttt{kstate\_emb}) and can be directly compared with the main grey-box results in Table~\ref{abres} (with \texttt{kstate\_emb}). The comparison reveals a stark and consistent performance collapse across all datasets, target models, and defense scenarios when the \texttt{kstate\_emb} is withheld. For instance, when attacking the undefended KSCD model on the Math1 dataset, the MIAttacker's AUC drops precipitously from a near-perfect 1.0000 (with \texttt{kstate\_emb}) to a mere 0.6335 (without). This dramatic degradation is not an isolated case but a systematic pattern observed throughout the experiments. Detailed results can be founded at \textbf{Appendix B.2}.

The findings from this study are conclusive. The massive performance gap between the full grey-box models and their ablated, ``knowledge-blind'' counterparts provides decisive evidence that the \texttt{kstate\_emb} is not just a supplementary feature, but the dominant source of the privacy leakage we exploit. It confirms that the learner profile itself constitutes a severe, standalone vulnerability, and access to this internal state information is what elevates our P-MIA from a standard black-box attack to a significantly more potent threat.

\subsection{Auditing Machine Unlearning Defenses with P-MIA}
To demonstrate the utility of P-MIA as an advanced auditing tool, we evaluate its ability to quantify the effectiveness of various machine unlearning defense strategies. The very premise of this grey-box audit rests on the attacker's ability to reliably obtain the \texttt{kstate\_emb}, as established in our threat model and detailed in Section~4.2. To ground this audit in a realistic scenario, we first empirically validated this extraction process.

We implemented and tested two automated methods for reverse-engineering the \texttt{kstate\_emb} from visual radar charts: a computer vision algorithm based on Canny edge detection and a multi-modal LLM (GPT-4o-mini). A representative result is shown in Figure~\ref{vs_radar}. Our vision-based method demonstrates exceptionally high fidelity, achieving a mean absolute error (MAE) between 0.01--0.03 compared to the ground-truth values. The LLM also provides a viable, albeit slightly less precise, alternative. This validation confirms that attackers can reliably obtain the necessary grey-box features, making our audit scenario highly plausible. More comprehensive results for this experiment are available in \textbf{Appendix B.3}.

Figure~\ref{vs} audits the effectiveness of unlearning defenses using our P-MIA framework. The results confirm two main findings. First, P-MIA is a sensitive audit tool, clearly distinguishing the highly vulnerable undefended models (``No Defense'') from the effectively protected retrained models (``Ideal (Retrain)''). Second, and most importantly, the audit reveals that current approximate unlearning methods provide insufficient protection for CDMs. Even after applying defenses like Amnesiac, L-CODEC, or SSD, the attack AUC on most models remains substantially above the 0.5 random-guessing baseline. This demonstrates that significant privacy risks persist, highlighting the inadequacy of general-purpose unlearning algorithms for the specific architecture of CDMs and underscoring the need for more robust, domain-specific solutions.
\section{Conclusion}
\label{sec:conclusion}

In this paper, we conducted the first systematic investigation of MIA against modern CDMs. We introduced P-MIA, a novel grey-box threat model that exploits the internal knowledge state vectors exposed by platforms' explainability features, such as learner profiles. Our experiments demonstrated that this information, which can be reliably reverse-engineered from visual charts, allows our grey-box attack to significantly outperform standard black-box baselines. Furthermore, we showcased P-MIA's utility as a high-fidelity auditing tool, successfully quantifying the limitations of various machine unlearning defenses on CDMs. Our audit revealed that current defenses are largely insufficient, leaving significant privacy risks unaddressed. This work highlights a critical trade-off between model explainability and user privacy in educational AI, underscoring the need for domain-specific defenses that can mitigate such grey-box threats without sacrificing explainability.
\bibliography{aaai2026}

\begin{thebibliography}{30}
\providecommand{\natexlab}[1]{#1}

\bibitem[{Bourtoule et~al.(2021)Bourtoule, Chandrasekaran, Choquette-Choo, Jia, Travers, Zhang, Lie, and Papernot}]{mu_sisa}
Bourtoule, L.; Chandrasekaran, V.; Choquette-Choo, C.~A.; Jia, H.; Travers, A.; Zhang, B.; Lie, D.; and Papernot, N. 2021.
\newblock Machine unlearning.
\newblock In \emph{2021 IEEE symposium on security and privacy (SP)}, 141--159. IEEE.

\bibitem[{Brzezi{\'n}ska(2020)}]{irt2}
Brzezi{\'n}ska, J. 2020.
\newblock Item response theory models in the measurement theory.
\newblock \emph{Communications in Statistics-Simulation and Computation}, 49(12): 3299--3313.

\bibitem[{Choi and Na(2025)}]{mu_fisher1}
Choi, D.; and Na, D. 2025.
\newblock Distribution-Level Feature Distancing for Machine Unlearning: Towards a Better Trade-off Between Model Utility and Forgetting.
\newblock In \emph{Proceedings of the AAAI Conference on Artificial Intelligence}, volume~39, 2536--2544.

\bibitem[{Detection(2009)}]{canny}
Detection, C.~E. 2009.
\newblock Canny Edge Detection.
\newblock \emph{Differences}, 180: 200.

\bibitem[{Foster, Schoepf, and Brintrup(2024)}]{mu_ssd}
Foster, J.; Schoepf, S.; and Brintrup, A. 2024.
\newblock Fast machine unlearning without retraining through selective synaptic dampening.
\newblock In \emph{Proceedings of the AAAI conference on artificial intelligence}, volume~38, 12043--12051.

\bibitem[{Gao et~al.(2021)Gao, Liu, Huang, Yin, Bi, Wang, Ma, Wang, and Su}]{graphcd_rcd}
Gao, W.; Liu, Q.; Huang, Z.; Yin, Y.; Bi, H.; Wang, M.-C.; Ma, J.; Wang, S.; and Su, Y. 2021.
\newblock RCD: Relation map driven cognitive diagnosis for intelligent education systems.
\newblock In \emph{Proceedings of the 44th international ACM SIGIR conference on research and development in information retrieval}, 501--510.

\bibitem[{Graves, Nagisetty, and Ganesh(2021)}]{mu_ga1}
Graves, L.; Nagisetty, V.; and Ganesh, V. 2021.
\newblock Amnesiac machine learning.
\newblock In \emph{Proceedings of the AAAI Conference on Artificial Intelligence}, volume~35, 11516--11524.

\bibitem[{Guo et~al.(2019)Guo, Goldstein, Hannun, and Van Der~Maaten}]{mu_hessian2}
Guo, C.; Goldstein, T.; Hannun, A.; and Van Der~Maaten, L. 2019.
\newblock Certified data removal from machine learning models.
\newblock \emph{arXiv preprint arXiv:1911.03030}.

\bibitem[{Hu et~al.(2022)Hu, Salcic, Sun, Dobbie, Yu, and Zhang}]{survey_mia}
Hu, H.; Salcic, Z.; Sun, L.; Dobbie, G.; Yu, P.~S.; and Zhang, X. 2022.
\newblock Membership inference attacks on machine learning: A survey.
\newblock \emph{ACM Computing Surveys (CSUR)}, 54(11s): 1--37.

\bibitem[{Johns, Mahadevan, and Woolf(2006)}]{irt1}
Johns, J.; Mahadevan, S.; and Woolf, B. 2006.
\newblock Estimating student proficiency using an item response theory model.
\newblock In \emph{International conference on intelligent tutoring systems}, 473--480. Springer.

\bibitem[{Li et~al.(2022)Li, Hu, Shuai, Yang, Zhang, Dai, and Xiong}]{neuralncd}
Li, G.; Hu, Y.; Shuai, J.; Yang, T.; Zhang, Y.; Dai, S.; and Xiong, N. 2022.
\newblock NeuralNCD: A neural network cognitive diagnosis model based on multi-dimensional features.
\newblock \emph{Applied Sciences}, 12(19): 9806.

\bibitem[{Liu et~al.(2022)Liu, Xu, Yuan, Wang, and Li}]{mu_fisher2}
Liu, Y.; Xu, L.; Yuan, X.; Wang, C.; and Li, B. 2022.
\newblock The right to be forgotten in federated learning: An efficient realization with rapid retraining.
\newblock In \emph{IEEE INFOCOM 2022-IEEE conference on computer communications}, 1749--1758. IEEE.

\bibitem[{Ma et~al.(2022)Ma, Li, Wu, Zhang, Cao, Zhang, and Zhao}]{kscd}
Ma, H.; Li, M.; Wu, L.; Zhang, H.; Cao, Y.; Zhang, X.; and Zhao, X. 2022.
\newblock Knowledge-sensed cognitive diagnosis for intelligent education platforms.
\newblock In \emph{Proceedings of the 31st ACM international conference on information \& knowledge management}, 1451--1460.

\bibitem[{Mantelero(2013)}]{gdpr}
Mantelero, A. 2013.
\newblock The eu proposal for a general data protection regulation and the roots of the ‘right to be forgotten’.
\newblock \emph{Computer Law \& Security Review}, 29(3): 229--235.

\bibitem[{Mehta et~al.(2022)Mehta, Pal, Singh, and Ravi}]{mu_hessian1}
Mehta, R.; Pal, S.; Singh, V.; and Ravi, S.~N. 2022.
\newblock Deep unlearning via randomized conditionally independent hessians.
\newblock In \emph{Proceedings of the IEEE/CVF Conference on Computer Vision and Pattern Recognition}, 10422--10431.

\bibitem[{Nasr, Shokri, and Houmansadr(2019)}]{mia_bwbox}
Nasr, M.; Shokri, R.; and Houmansadr, A. 2019.
\newblock Comprehensive privacy analysis of deep learning: Passive and active white-box inference attacks against centralized and federated learning.
\newblock In \emph{2019 IEEE symposium on security and privacy (SP)}, 739--753. IEEE.

\bibitem[{Nguyen et~al.(2022)Nguyen, Huynh, Ren, Nguyen, Liew, Yin, and Nguyen}]{survey_mu2}
Nguyen, T.~T.; Huynh, T.~T.; Ren, Z.; Nguyen, P.~L.; Liew, A. W.-C.; Yin, H.; and Nguyen, Q. V.~H. 2022.
\newblock A survey of machine unlearning.
\newblock \emph{arXiv preprint arXiv:2209.02299}.

\bibitem[{Shokri et~al.(2017)Shokri, Stronati, Song, and Shmatikov}]{mia_bc1}
Shokri, R.; Stronati, M.; Song, C.; and Shmatikov, V. 2017.
\newblock Membership inference attacks against machine learning models.
\newblock In \emph{2017 IEEE symposium on security and privacy (SP)}, 3--18. IEEE.

\bibitem[{Tarun et~al.(2023)Tarun, Chundawat, Mandal, and Kankanhalli}]{mu_ga2}
Tarun, A.~K.; Chundawat, V.~S.; Mandal, M.; and Kankanhalli, M. 2023.
\newblock Fast yet effective machine unlearning.
\newblock \emph{IEEE Transactions on Neural Networks and Learning Systems}, 35(9): 13046--13055.

\bibitem[{Wang et~al.(2024{\natexlab{a}})Wang, Gao, Liu, Li, Zhao, Zhang, Huang, Zhu, Wang, Tong et~al.}]{survey_cd}
Wang, F.; Gao, W.; Liu, Q.; Li, J.; Zhao, G.; Zhang, Z.; Huang, Z.; Zhu, M.; Wang, S.; Tong, W.; et~al. 2024{\natexlab{a}}.
\newblock A survey of models for cognitive diagnosis: New developments and future directions.
\newblock \emph{arXiv preprint arXiv:2407.05458}.

\bibitem[{Wang et~al.(2020)Wang, Liu, Chen, Huang, Chen, Yin, Huang, and Wang}]{neuralcd}
Wang, F.; Liu, Q.; Chen, E.; Huang, Z.; Chen, Y.; Yin, Y.; Huang, Z.; and Wang, S. 2020.
\newblock Neural cognitive diagnosis for intelligent education systems.
\newblock In \emph{Proceedings of the AAAI conference on artificial intelligence}, volume~34, 6153--6161.

\bibitem[{Wang et~al.(2022)Wang, Liu, Chen, Huang, Yin, Wang, and Su}]{kancd}
Wang, F.; Liu, Q.; Chen, E.; Huang, Z.; Yin, Y.; Wang, S.; and Su, Y. 2022.
\newblock NeuralCD: a general framework for cognitive diagnosis.
\newblock \emph{IEEE Transactions on Knowledge and Data Engineering}, 35(8): 8312--8327.

\bibitem[{Wang et~al.(2024{\natexlab{b}})Wang, Tian, Zhang, and Yu}]{survey_mu1}
Wang, W.; Tian, Z.; Zhang, C.; and Yu, S. 2024{\natexlab{b}}.
\newblock Machine unlearning: A comprehensive survey.
\newblock \emph{arXiv preprint arXiv:2405.07406}.

\bibitem[{Wen et~al.(2024)Wen, Li, Backes, and Zhang}]{mia_icl}
Wen, R.; Li, Z.; Backes, M.; and Zhang, Y. 2024.
\newblock Membership inference attacks against in-context learning.
\newblock In \emph{Proceedings of the 2024 on ACM SIGSAC Conference on Computer and Communications Security}, 3481--3495.

\bibitem[{Yeom et~al.(2018)Yeom, Giacomelli, Fredrikson, and Jha}]{mia_overfitting}
Yeom, S.; Giacomelli, I.; Fredrikson, M.; and Jha, S. 2018.
\newblock Privacy risk in machine learning: Analyzing the connection to overfitting.
\newblock In \emph{2018 IEEE 31st computer security foundations symposium (CSF)}, 268--282. IEEE.

\bibitem[{Zanella-Beguelin et~al.(2021)Zanella-Beguelin, Tople, Paverd, and K{\"o}pf}]{greybox}
Zanella-Beguelin, S.; Tople, S.; Paverd, A.; and K{\"o}pf, B. 2021.
\newblock Grey-box extraction of natural language models.
\newblock In \emph{International Conference on Machine Learning}, 12278--12286. PMLR.

\bibitem[{Zhang, Zhao, and Zhang(2024)}]{mia_xgboost1}
Zhang, G.; Zhao, P.; and Zhang, A. 2024.
\newblock LocMIA: Membership Inference Attacks Against Aggregated Location Data.
\newblock In \emph{Privacy Preservation in Distributed Systems: Algorithms and Applications}, 3--24. Springer.

\bibitem[{Zhao et~al.(2025)Zhao, Huang, Cheng, Zhuang, Mao, Li, Wang, and Chen}]{graphcd_dmccdm}
Zhao, G.; Huang, Z.; Cheng, C.; Zhuang, Y.; Mao, Q.; Li, X.; Wang, S.; and Chen, E. 2025.
\newblock Multi-Perspective Consolidation Enhanced Cognitive Diagnosis via Conditional Diffusion Model.
\newblock In \emph{Proceedings of the AAAI Conference on Artificial Intelligence}, volume~39, 1174--1182.

\bibitem[{Zhao et~al.(2023)Zhao, Zhou, Li, Tang, Wang, Hou, Min, Zhang, Zhang, Dong et~al.}]{llm}
Zhao, W.~X.; Zhou, K.; Li, J.; Tang, T.; Wang, X.; Hou, Y.; Min, Y.; Zhang, B.; Zhang, J.; Dong, Z.; et~al. 2023.
\newblock A survey of large language models.
\newblock \emph{arXiv preprint arXiv:2303.18223}, 1(2).

\bibitem[{Zhou et~al.(2021)Zhou, Liu, Wu, Wang, Huang, Tong, Xiong, Chen, and Ma}]{ecd}
Zhou, Y.; Liu, Q.; Wu, J.; Wang, F.; Huang, Z.; Tong, W.; Xiong, H.; Chen, E.; and Ma, J. 2021.
\newblock Modeling context-aware features for cognitive diagnosis in student learning.
\newblock In \emph{Proceedings of the 27th ACM SIGKDD conference on knowledge discovery \& data mining}, 2420--2428.

\end{thebibliography}
\appendix
\section{Supplementary Information on Experimental Settings}
\subsection{Datasets}
A detailed description of the datasets used in this paper is as follows:

\begin{itemize}
    \item \textbf{Math1 \& Math2:} These two datasets are derived from two large-scale high school final examinations, representing more complex and realistic diagnostic scenarios.
    \begin{itemize}
        \item The \textbf{Math1} dataset contains 84,180 response records from 4,209 students on 20 questions, which are associated with 11 KCs in the domain of algebra.
        \item The \textbf{Math2} dataset contains 78,220 response records from 3,911 students on 20 questions, covering 16 KCs spanning both algebra and geometry.
    \end{itemize}
    A key characteristic of these datasets is that the exercises often require the simultaneous mastery of multiple KCs, making them ideal for evaluating model performance in complex educational assessment settings.

    \item \textbf{FrcSub:} This is a classic benchmark dataset in the CD field, originally introduced by Tatsuoka. It contains the response records of 536 middle school students on 20 fraction subtraction problems. The value of this dataset lies in its fine-grained Q-matrix, which defines the 8 underlying cognitive skills (e.g., simplifying fractions, converting whole numbers to fractions) required to perform fraction subtraction. Due to its focused domain and well-defined skill set, FrcSub is widely used to evaluate the fine-grained diagnostic capabilities of CD models.
\end{itemize}
\subsection{Baselines}
This section provides detailed descriptions of the target models and defense mechanisms used in our experiments.

\subsubsection{Target Cognitive Diagnosis Models}
We selected three representative deep learning-based CDMs that are widely recognized in the field.
\paragraph{NeuralCD~\cite{kancd}}
NeuralCD is a foundational and general framework for cognitive diagnosis that replaces traditional, handcrafted interaction functions (like those in IRT) with a MLP. It learns the complex, non-linear interactions between student proficiency and exercise attributes directly from data. To maintain interpretability, a key requirement for CDMs, NeuralCD introduces a \textit{monotonicity assumption}, ensuring that a student's predicted performance on an exercise does not decrease as their mastery of a required knowledge concept increases. It serves as the basis for many modern CDMs.

\paragraph{KSCD~\cite{kscd}}
Knowledge-sensed CD (KSCD) is an advanced CDM designed to learn the intrinsic relationships among different knowledge concepts. It projects students, exercises, and knowledge concepts into a unified embedding space. A student's knowledge state vector is then generated via a multiplication product between the student's embedding and the knowledge concept embeddings. This ``knowledge-sensed'' approach allows the model to infer student proficiency even on concepts they have not directly practiced, addressing data sparsity issues.

\paragraph{KaNCD~\cite{kancd}}
The knowledge-association aware NCD (KaNCD) is a specific extension proposed within the NeuralCD framework to tackle the knowledge coverage problem. When a student's response records are sparse, their proficiency on many KCs cannot be directly estimated. KaNCD addresses this by explicitly modeling the associations between knowledge concepts, allowing the model to infer proficiency on ``unseen'' KCs based on the mastery of related, ``seen'' KCs.

\subsubsection{Machine Unlearning Defenses}
We evaluate P-MIA's auditing capabilities against three state-of-the-art approximate MU approaches, each representing a different algorithmic family.

\paragraph{Gradient Ascent-based (Amnesiac)}
This approach is based on the intuitive principle of reversing the training process for the data to be forgotten~\cite{mu_ga1}. Instead of minimizing the loss on the forget set $\mathcal{D}_f$ via gradient descent, this method performs gradient \textit{ascent} for a small number of steps. This update pushes the model's parameters away from the region that correctly classifies the forget data, effectively making the model forget those samples. It is computationally efficient as it only requires a few forward and backward passes on the small forget set.

\paragraph{Hessian-based (L-CODEC)}
Hessian-based unlearning methods~\cite{mu_hessian1} offer a more principled approach by approximating a single Newton update step to remove a sample's influence on the model parameters. The core idea relies on the Hessian matrix (the second-order derivative of the loss function), which captures the curvature of the loss landscape. As calculating the full Hessian is computationally intractable for deep models, our implementation, in line with modern efficient approaches, uses the Hutchinson's method to approximate the Hessian's diagonal. This is achieved by computing Hessian-vector products (HVPs), as shown in the provided code snippet. The estimated influence of the forget set is then subtracted from the original model's weights.

\begin{figure*}[ht]
\centering
\includegraphics[width=0.85\textwidth]{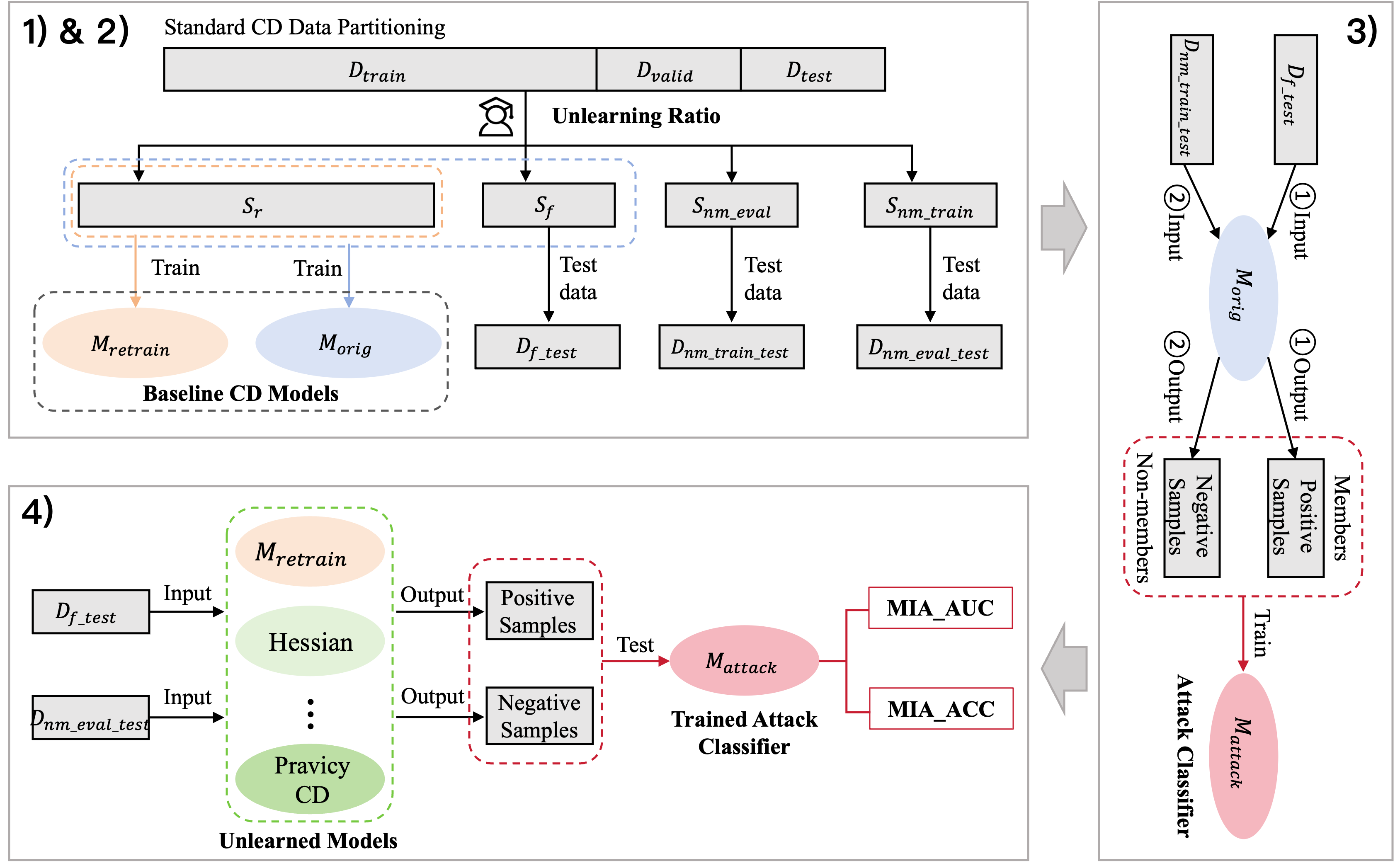} 
\caption{The overall pipeline of MIA Protocol.}
\label{pipe_mia}
\end{figure*}
\begin{table*}[ht]
\centering
\renewcommand{\arraystretch}{1.5}
\resizebox{\textwidth}{!}{%
\begin{tabular}{cc|cccccc|cccccc|cccccc}
\hline
\multicolumn{2}{c|}{\textbf{Dataset}} &
  \multicolumn{6}{c|}{\textbf{Math1}} &
  \multicolumn{6}{c|}{\textbf{Math2}} &
  \multicolumn{6}{c}{\textbf{Frcsub}} \\ \hline
\multicolumn{2}{c|}{\textbf{Target w/ Defenses}} &
  \multicolumn{2}{c}{\textbf{XGBoost}} &
  \multicolumn{2}{c}{\textbf{DCA}} &
  \multicolumn{2}{c|}{\textbf{MIAttacker}} &
  \multicolumn{2}{c}{\textbf{XGBoost}} &
  \multicolumn{2}{c}{\textbf{DCA}} &
  \multicolumn{2}{c|}{\textbf{MIAttacker}} &
  \multicolumn{2}{c}{\textbf{XGboost}} &
  \multicolumn{2}{c}{\textbf{DCA}} &
  \multicolumn{2}{c}{\textbf{MIAttacker}} \\ \hline
\textbf{Target Model} &
  \textbf{Defenses} &
  ACC &
  AUC &
  ACC &
  AUC &
  ACC &
  AUC &
  ACC &
  AUC &
  ACC &
  AUC &
  ACC &
  AUC &
  ACC &
  AUC &
  ACC &
  AUC &
  ACC &
  AUC \\ \hline
\multirow{4}{*}{NeuralCD} &
  * &
  0.8125 &
  0.8690 &
  0.9953 &
  0.9996 &
  0.9871 &
  0.9993 &
  0.7886 &
  0.8471 &
  0.9984 &
  1.0000 &
  0.9918 &
  0.9995 &
  0.7407 &
  0.8136 &
  0.9912 &
  1.0000 &
  0.9868 &
  0.9995 \\
 &
  Amnesiac &
  0.7644 &
  0.8333 &
  1.0000 &
  1.0000 &
  0.9907 &
  0.9993 &
  0.7980 &
  0.8545 &
  0.9984 &
  1.0000 &
  0.9918 &
  0.9995 &
  0.6989 &
  0.7490 &
  0.9912 &
  1.0000 &
  0.9868 &
  0.9995 \\
 &
  L-CODEC &
  0.7263 &
  0.8101 &
  0.9938 &
  0.9978 &
  0.9906 &
  0.9978 &
  0.6247 &
  0.7226 &
  0.9973 &
  1.0000 &
  0.9918 &
  0.9948 &
  0.5670 &
  0.6242 &
  0.9341 &
  0.9928 &
  0.9363 &
  0.9917 \\
 &
  SSD &
  \textit{0.6394} &
  \textit{0.6734} &
  \textbf{0.9723} &
  \textbf{0.9967} &
  0.9466 &
  0.9927 &
  \textit{0.6020} &
  \textit{0.5737} &
  \textbf{0.9808} &
  \textbf{0.9997} &
  0.9651 &
  0.9879 &
  \textit{0.5495} &
  \textit{0.6127} &
  0.8703 &
  \textbf{0.9988} &
  \textbf{0.8967} &
  0.9735 \\ \hline
\multirow{4}{*}{KSCD} &
  * &
  0.9041 &
  0.9522 &
  0.9973 &
  1.0000 &
  0.9887 &
  0.9985 &
  0.7996 &
  0.8537 &
  0.9769 &
  0.9931 &
  0.9588 &
  0.9831 &
  0.6989 &
  0.7860 &
  0.9538 &
  0.9710 &
  0.9385 &
  0.9557 \\
 &
  Amnesiac &
  0.8530 &
  0.8945 &
  0.9953 &
  1.0000 &
  0.9930 &
  1.0000 &
  0.8020 &
  0.8535 &
  0.9765 &
  0.9926 &
  0.9580 &
  0.9826 &
  0.6989 &
  0.7794 &
  0.9560 &
  0.9710 &
  0.9385 &
  0.9559 \\
 &
  L-CODEC &
  0.8698 &
  0.9230 &
  0.9747 &
  0.9775 &
  0.9712 &
  0.9756 &
  0.6976 &
  0.7539 &
  0.9412 &
  0.9556 &
  0.9239 &
  0.9415 &
  0.5319 &
  0.5465 &
  0.7429 &
  0.6607 &
  0.7429 &
  0.7534 \\
 &
  SSD &
  \textit{0.7283} &
  \textit{0.7668} &
  \textbf{0.7567} &
  0.7213 &
  0.7501 &
  \textbf{0.7668} &
  \textit{0.6294} &
  \textit{0.6211} &
  0.7239 &
  0.6142 &
  \textbf{0.7267} &
  \textbf{0.6406} &
  \textit{0.5165} &
  \textit{0.5105} &
  \textbf{0.6000} &
  \textbf{0.6295} &
  0.5868 &
  0.5829 \\ \hline
\multirow{4}{*}{KaNCD} &
  * &
  0.7025 &
  0.7731 &
  0.8955 &
  0.9346 &
  0.8951 &
  0.9328 &
  0.5686 &
  0.6030 &
  0.7522 &
  0.7899 &
  0.7522 &
  0.7868 &
  0.6989 &
  0.7604 &
  0.8813 &
  0.9558 &
  0.8132 &
  0.8995 \\
 &
  Amnesiac &
  0.6854 &
  0.7487 &
  0.8912 &
  0.9348 &
  0.8951 &
  0.9346 &
  0.5678 &
  0.5992 &
  0.7490 &
  0.7884 &
  0.7510 &
  0.7856 &
  0.6725 &
  0.7328 &
  0.8769 &
  0.9522 &
  0.8110 &
  0.8916 \\
 &
  L-CODEC &
  0.5556 &
  0.6084 &
  0.8834 &
  0.9464 &
  0.8838 &
  0.9322 &
  0.4655 &
  0.4481 &
  0.5671 &
  0.7796 &
  0.8839 &
  0.9490 &
  \textit{0.4835} &
  \textit{0.4791} &
  0.6176 &
  0.6721 &
  \textbf{0.6593} &
  \textbf{0.6804} \\
 &
  SSD &
  \textit{0.5708} &
  \textit{0.5864} &
  \textbf{0.6620} &
  \textbf{0.6679} &
  0.6604 &
  0.5215 &
  \textit{0.5031} &
  \textit{0.5112} &
  \textbf{0.5973} &
  \textbf{0.4937} &
  0.5922 &
  0.4804 &
  0.4505 &
  0.3798 &
  0.6352 &
  0.6672 &
  0.5934 &
  0.6302 \\ \hline
\end{tabular}%
}
\caption{MIA results on all datasets for the 10\% forgetting ratio scenario. We compare the black-box (XGBoost) and grey-box (DCA, MIAttacker) attacks. `*' denotes no defense. \textit{Italicized} values mark the most effective defense against the black-box attack, while \textbf{bold} values highlight the best performance for each of our grey-box attackers within a model block.}
\label{mainres_01}
\end{table*}
\begin{table*}[ht]
\centering
\renewcommand{\arraystretch}{1.5}
\resizebox{\textwidth}{!}{%
\begin{tabular}{cc|cccccc|cccccc|cccccc}
\hline
\multicolumn{2}{c|}{\textbf{Dataset}} &
  \multicolumn{6}{c|}{\textbf{Math1}} &
  \multicolumn{6}{c|}{\textbf{Math2}} &
  \multicolumn{6}{c}{\textbf{Frcsub}} \\ \hline
\multicolumn{2}{c|}{\textbf{Target w/ Defenses}} &
  \multicolumn{2}{c}{\textbf{XGBoost}} &
  \multicolumn{2}{c}{\textbf{DCA}} &
  \multicolumn{2}{c|}{\textbf{MIAttacker}} &
  \multicolumn{2}{c}{\textbf{XGBoost}} &
  \multicolumn{2}{c}{\textbf{DCA}} &
  \multicolumn{2}{c|}{\textbf{MIAttacker}} &
  \multicolumn{2}{c}{\textbf{XGboost}} &
  \multicolumn{2}{c}{\textbf{DCA}} &
  \multicolumn{2}{c}{\textbf{MIAttacker}} \\ \hline
\textbf{Target Model} &
  \textbf{Defenses} &
  ACC &
  AUC &
  ACC &
  AUC &
  ACC &
  AUC &
  ACC &
  AUC &
  ACC &
  AUC &
  ACC &
  AUC &
  ACC &
  AUC &
  ACC &
  AUC &
  ACC &
  AUC \\ \hline
\multirow{4}{*}{NeuralCD} &
  * &
  0.6811 &
  0.7302 &
  0.9882 &
  1.0000 &
  0.9488 &
  0.9941 &
  0.6667 &
  0.7255 &
  1.0000 &
  1.0000 &
  0.9719 &
  0.9988 &
  0.7143 &
  0.7943 &
  1.0000 &
  1.0000 &
  0.7857 &
  0.9636 \\
 &
  Amnesiac &
  0.6850 &
  0.7387 &
  0.9882 &
  1.0000 &
  0.9488 &
  0.9941 &
  0.5944 &
  0.6689 &
  1.0000 &
  1.0000 &
  0.9719 &
  0.9988 &
  0.6905 &
  0.7818 &
  1.0000 &
  1.0000 &
  0.7857 &
  0.9636 \\
 &
  L-CODEC &
  0.6260 &
  0.6233 &
  0.9449 &
  0.9794 &
  0.9409 &
  0.9504 &
  0.5462 &
  0.5979 &
  0.9960 &
  1.0000 &
  0.9759 &
  0.9972 &
  0.6667 &
  0.7875 &
  0.8095 &
  0.7955 &
  0.8571 &
  0.8614 \\
 &
  SSD &
  \textit{0.5236} &
  \textit{0.5191} &
  \textbf{0.9528} &
  0.9787 &
  0.9449 &
  \textbf{0.9945} &
  \textit{0.5341} &
  \textit{0.5379} &
  \textbf{0.9839} &
  \textbf{0.9996} &
  0.9478 &
  0.9874 &
  \textit{0.5238} &
  \textit{0.5580} &
  \textbf{0.8095} &
  0.7455 &
  0.7619 &
  \textbf{0.8000} \\ \hline
\multirow{4}{*}{KSCD} &
  * &
  0.7756 &
  0.8180 &
  0.9803 &
  1.0000 &
  0.9646 &
  0.9996 &
  0.7470 &
  0.8162 &
  0.9357 &
  0.9525 &
  0.9036 &
  0.9595 &
  0.7143 &
  0.8045 &
  0.8571 &
  0.9045 &
  0.7143 &
  0.9068 \\
 &
  Amnesiac &
  0.7677 &
  0.8042 &
  0.9882 &
  1.0000 &
  0.9646 &
  0.9996 &
  0.7229 &
  0.7868 &
  0.9357 &
  0.9533 &
  0.9036 &
  0.9591 &
  0.6905 &
  0.7898 &
  0.8571 &
  0.9045 &
  0.7143 &
  0.9045 \\
 &
  L-CODEC &
  0.7126 &
  0.7460 &
  0.7441 &
  0.6930 &
  0.7205 &
  0.6468 &
  0.6104 &
  0.6898 &
  0.7309 &
  0.6532 &
  0.7149 &
  0.5286 &
  \textit{0.5000} &
  \textit{0.4830} &
  0.4286 &
  0.2682 &
  \textbf{0.5238} &
  \textbf{0.5705} \\
 &
  SSD &
  \textit{0.6024} &
  \textit{0.6109} &
  0.6457 &
  \textbf{0.7769} &
  \textbf{0.6969} &
  0.7659 &
  \textit{0.4739} &
  \textit{0.4618} &
  \textbf{0.5904} &
  0.4517 &
  0.5502 &
  \textbf{0.5359} &
  0.4048 &
  0.2705 &
  0.4286 &
  0.5500 &
  0.5238 &
  0.3795 \\ \hline
\multirow{4}{*}{KaNCD} &
  * &
  0.6260 &
  0.6574 &
  0.9331 &
  0.9526 &
  0.9331 &
  0.9498 &
  0.5582 &
  0.5647 &
  0.7028 &
  0.7746 &
  0.6867 &
  0.7630 &
  0.7381 &
  0.8511 &
  0.5714 &
  0.8250 &
  0.4762 &
  0.6364 \\
 &
  Amnesiac &
  0.6024 &
  0.6222 &
  0.9331 &
  0.9522 &
  0.9331 &
  0.9495 &
  0.5422 &
  0.5540 &
  0.7028 &
  0.7742 &
  0.6867 &
  0.7627 &
  0.7381 &
  0.8511 &
  0.5714 &
  0.8250 &
  0.4762 &
  0.6364 \\
 &
  L-CODEC &
  \textit{0.5197} &
  \textit{0.5256} &
  0.8071 &
  0.9789 &
  \textbf{0.8622} &
  \textbf{0.9940} &
  0.5221 &
  0.4276 &
  0.4137 &
  0.0235 &
  0.7952 &
  0.7605 &
  \textit{0.5238} &
  \textit{0.5205} &
  0.2143 &
  0.1659 &
  \textbf{0.4524} &
  \textbf{0.3705} \\
 &
  SSD &
  0.5709 &
  0.5273 &
  0.6142 &
  0.4268 &
  0.5945 &
  0.4909 &
  \textit{0.4739} &
  \textit{0.4763} &
  \textbf{0.4900} &
  \textbf{0.6479} &
  0.2851 &
  0.3045 &
  0.5476 &
  0.4648 &
  0.2619 &
  0.2477 &
  0.5238 &
  0.4432 \\ \hline
\end{tabular}%
}
\caption{MIA results on all datasets for the 1\% forgetting ratio scenario. We compare the black-box (XGBoost) and grey-box (DCA, MIAttacker) attacks. `*' denotes no defense. \textit{Italicized} values mark the most effective defense against the black-box attack, while \textbf{bold} values highlight the best performance for each of our grey-box attackers within a model block.}
\label{mainres_001}
\end{table*}
\paragraph{FIM-based (SSD)}
Selective synaptic dampening (SSD)~\cite{mu_ssd} is a method that leverages the FIM to estimate the importance of each model parameter with respect to the retain set ($\mathcal{D}_r$) and the forget set ($\mathcal{D}_f$). The FIM approximates the Hessian and indicates how sensitive the model's output is to changes in its parameters. SSD calculates two FIMs and then dampens (i.e., penalizes or regularizes) the parameters that were identified as highly important for the forget set, thereby selectively erasing the information associated with $\mathcal{D}_f$ while preserving the knowledge from $\mathcal{D}_r$.
\subsection{MIA Protocol for Auditing Unlearning Efficacy}
To rigorously quantify the effectiveness of unlearning defenses, we designed a comprehensive MIA protocol, visually summarized in Figure~\ref{pipe_mia}. The entire process is built upon a student-level data partitioning strategy to respect the user-centric nature of cognitive diagnosis data. The protocol unfolds in four key steps:

\begin{enumerate}[label=\textbf{Step \arabic*:}, leftmargin=*]
    \item \textbf{Data Partitioning.} 
    We begin with a standard dataset split into $\mathcal{D}_{train}$, $\mathcal{D}_{valid}$, and $\mathcal{D}_{test}$. From the student population, we then define four disjoint sets based on a predefined unlearning ratio: Retain Students ($\mathcal{S}_r$), Forget Students ($\mathcal{S}_f$), Non-member Train Students ($\mathcal{S}_{nm\_train}$), and Non-member Eval Students ($\mathcal{S}_{nm\_eval}$). These student-level sets are used to create the corresponding data subsets (e.g., $\mathcal{D}_{f\_test}$ contains test data from students in $\mathcal{S}_f$).

    \item \textbf{Baseline CD Model Training.}
    Two baseline models are trained to establish the upper and lower bounds for privacy leakage. The \textbf{Original Model ($M_{orig}$)} is trained on data from both $\mathcal{S}_r$ and $\mathcal{S}_f$, representing a vulnerable deployed system. The \textbf{Retrained Model ($M_{retrain}$)} is trained only on data from $\mathcal{S}_r$, serving as the gold standard for perfect unlearning.

    \item \textbf{Attack Classifier Training.}
    We train the attack model ($M_{attack}$) to learn the behavioral differences between members and non-members. To generate training data, we query the original model $M_{orig}$. The outputs from $M_{orig}$ on the test data of forget students ($\mathcal{D}_{f\_test}$) serve as \textit{Positive Samples (Members)}, while outputs on the test data of non-member train students ($\mathcal{D}_{nm\_train\_test}$) serve as \textit{Negative Samples (Non-members)}. $M_{attack}$ is trained on these labeled samples.

    \item \textbf{Unlearning Efficacy Evaluation.}
    Finally, we use the trained $M_{attack}$ to audit the privacy of various unlearned models (including $M_{retrain}$ and other approximate methods like Hessian-based unlearning). For each model, we generate test samples for the attacker by querying it on the forget set ($\mathcal{D}_{f\_test}$) for positive samples and the non-member evaluation set ($\mathcal{D}_{nm\_eval\_test}$) for negative samples. The resulting \textbf{MIA\_AUC} and \textbf{MIA\_ACC} scores quantify the unlearning efficacy: a score close to 0.5 indicates successful unlearning, while a score significantly higher suggests residual privacy leakage.
\end{enumerate}
\subsection{Implementation Details}
This section provides detailed implementation settings for our experiments to ensure full reproducibility.
\subsubsection{A.3.1 CD Model Architecture and Hyperparameter Tuning}
All our experiments are based on the neural CD architectures described in Appendix A.2. To ensure that our models achieve their optimal performance, we employed \textbf{Weights \& Biases (WanDB)}\footnote{https://wandb.ai/site/}, a leading platform for experiment tracking and hyperparameter optimization, to conduct a systematic hyperparameter search. The tuning process focused on key hyperparameters, including the embedding dimension, the hidden layer architecture of the MLP in the interaction module, learning rate, batch size, and dropout rates. By leveraging WandB's Bayesian search sweeps, we automatically explored the vast parameter space and selected an optimal configuration for each dataset based on the best performance on the validation set. All models, including the original ($M_{orig}$) and retrain ($M_{retrain}$) models, were subsequently trained using their respective optimal configurations. The complete scripts for our WanDB hyperparameter sweeps and the detailed experimental logs are available in our public code repository to ensure full reproducibility.

\subsubsection{A.3.2 Unlearning Hyperparameters Tuning}
The hyperparameter tuning for all unlearning (defense) algorithms was conducted on top of the optimal CD model configurations determined in the previous step. Unlike standard model training, the objective of unlearning is multi-faceted: it requires simultaneously maintaining high model utility on the retain set (high AUC/ACC) and ensuring unlearning completeness (MIA metrics close to the gold standard). The multi-objective nature of this task makes it less suitable for automated tuning platforms like WandB. Therefore, we developed custom tuning scripts to perform a grid search to identify the best-performing hyperparameter combination for each method on the validation set.

The specific search spaces for each method are detailed below:
\begin{itemize}
    
    \item \textbf{Amnesiac:} The unlearning learning rate was chosen from $\{10^{-5}, 5 \times 10^{-5}, 10^{-4}\}$, and the number of unlearning steps was chosen from $\{1, 3, 5\}$.
    
    \item \textbf{L-CODEC:} We tuned its two key parameters: the number of samples for Hessian estimation was searched in $\{10, 20, 40\}$, and the number of batches for Hessian-Vector Product (HVP) computation was selected from $\{1, 2\}$.
    
    \item \textbf{SSD:} The selection threshold $\alpha$ was searched in $\{1.3, 2.0, 2.5, 5.0\}$, and the unlearning strength $\lambda$ was searched in $\{0.1, 0.3, 0.5, 0.8\}$.
\end{itemize}
For each defense method, we performed an extensive hyperparameter search. The optimal configuration for each method was selected based on a single criterion: maximizing the unlearning efficacy. Specifically, we chose the set of hyperparameters that resulted in the attack model's performance (both MIA\_ACC and MIA\_AUC) being closest to 0.5, the random guess baseline, which emulates the security level of the ideally retrained model.

\begin{table*}[ht]
\centering
\renewcommand{\arraystretch}{1.2}
\resizebox{\textwidth}{!}{%
\begin{tabular}{cc|cccc|cccc|cccc}
\hline
\multicolumn{2}{c|}{\textbf{Dataset}} &
  \multicolumn{4}{c|}{\textbf{Math1}} &
  \multicolumn{4}{c|}{\textbf{Math2}} &
  \multicolumn{4}{c}{\textbf{Frcsub}} \\ \hline
\multicolumn{2}{c|}{\textbf{Target w/ Defenses}} &
  \multicolumn{2}{c}{\textbf{\begin{tabular}[c]{@{}c@{}}DCA w/o\\ kstate\_emb\end{tabular}}} &
  \multicolumn{2}{c|}{\textbf{\begin{tabular}[c]{@{}c@{}}MIAttacker w/o\\ kstate\_emb\end{tabular}}} &
  \multicolumn{2}{c}{\textbf{\begin{tabular}[c]{@{}c@{}}DCA w/o\\ kstate\_emb\end{tabular}}} &
  \multicolumn{2}{c|}{\textbf{\begin{tabular}[c]{@{}c@{}}MIAttacker w/o\\ kstate\_emb\end{tabular}}} &
  \multicolumn{2}{c}{\textbf{\begin{tabular}[c]{@{}c@{}}DCA w/o\\ kstate\_emb\end{tabular}}} &
  \multicolumn{2}{c}{\textbf{\begin{tabular}[c]{@{}c@{}}MIAttacker w/o\\ kstate\_emb\end{tabular}}} \\ \hline
\textbf{Target Model} &
  \textbf{Defenses} &
  $\text{ACC}_{\text{MIA}}$ &
  $\text{AUC}_{\text{MIA}}$ &
  $\text{ACC}_{\text{MIA}}$ &
  $\text{AUC}_{\text{MIA}}$ &
  $\text{ACC}_{\text{MIA}}$ &
  $\text{AUC}_{\text{MIA}}$ &
  $\text{ACC}_{\text{MIA}}$ &
  $\text{AUC}_{\text{MIA}}$ &
  $\text{ACC}_{\text{MIA}}$ &
  $\text{AUC}_{\text{MIA}}$ &
  $\text{ACC}_{\text{MIA}}$ &
  $\text{AUC}_{\text{MIA}}$ \\ \hline
\multirow{4}{*}{NeuralCD} &
  * &
  0.6912 &
  0.7575 &
  0.4990 &
  0.5513 &
  0.6165 &
  0.6936 &
  \textbf{0.6216} &
  \textbf{0.6323} &
  \textbf{0.6989} &
  \textbf{0.7894} &
  \textbf{0.6352} &
  \textbf{0.7386} \\
 &
  Amnesiac &
  \textbf{0.6990} &
  \textbf{0.7667} &
  \textbf{0.5310} &
  \textbf{0.5534} &
  \textbf{0.6263} &
  \textbf{0.7087} &
  0.6133 &
  0.6232 &
  0.6923 &
  0.7576 &
  \textbf{0.6352} &
  0.7165 \\
 &
  L-CODEC &
  0.5828 &
  0.6817 &
  0.4144 &
  0.4083 &
  0.4239 &
  0.4723 &
  0.4075 &
  0.3962 &
  0.5319 &
  0.5320 &
  0.5187 &
  0.5996 \\
 &
  SSD &
  0.5322 &
  0.5553 &
  0.4464 &
  0.4538 &
  0.5220 &
  0.5316 &
  0.5176 &
  0.5324 &
  0.5846 &
  0.5854 &
  0.5824 &
  0.6140 \\ \hline
\multirow{4}{*}{KSCD} &
  * &
  \textbf{0.7002} &
  \textbf{0.7820} &
  0.5871 &
  \textbf{0.6473} &
  0.6608 &
  0.7139 &
  \textbf{0.6176} &
  0.6128 &
  \textbf{0.6835} &
  0.7901 &
  0.6308 &
  0.7429 \\
 &
  Amnesiac &
  0.6982 &
  0.7790 &
  \textbf{0.5887} &
  0.6456 &
  \textbf{0.6663} &
  \textbf{0.7244} &
  0.6000 &
  \textbf{0.6196} &
  \textbf{0.6835} &
  \textbf{0.7910} &
  \textbf{0.6549} &
  \textbf{0.7435} \\
 &
  L-CODEC &
  0.6257 &
  0.6718 &
  0.3427 &
  0.3811 &
  0.4408 &
  0.4699 &
  0.3682 &
  0.2763 &
  0.5209 &
  0.5912 &
  0.5231 &
  0.5451 \\
 &
  SSD &
  0.5739 &
  0.6184 &
  0.5209 &
  0.5446 &
  0.5792 &
  0.5924 &
  0.5545 &
  0.5532 &
  0.4615 &
  0.4755 &
  0.5033 &
  0.5203 \\ \hline
\multirow{4}{*}{KaNCD} &
  * &
  0.6538 &
  \textbf{0.7015} &
  \textbf{0.6039} &
  \textbf{0.6400} &
  0.5753 &
  \textbf{0.6224} &
  \textbf{0.5392} &
  \textbf{0.5565} &
  0.6879 &
  0.7638 &
  \textbf{0.6725} &
  \textbf{0.7441} \\
 &
  Amnesiac &
  \textbf{0.6554} &
  0.6854 &
  0.5945 &
  0.6143 &
  \textbf{0.5757} &
  0.6152 &
  0.5345 &
  0.5492 &
  \textbf{0.6989} &
  \textbf{0.7754} &
  0.6615 &
  0.7424 \\
 &
  L-CODEC &
  0.4889 &
  0.4306 &
  0.3786 &
  0.3234 &
  0.3710 &
  0.3491 &
  0.2804 &
  0.2325 &
  0.3802 &
  0.3657 &
  0.4418 &
  0.4665 \\
 &
  SSD &
  0.5517 &
  0.5542 &
  0.5212 &
  0.5309 &
  0.5345 &
  0.5478 &
  0.5314 &
  0.5347 &
  0.3451 &
  0.2241 &
  0.3099 &
  0.3611 \\ \hline
\end{tabular}%
}
\caption{Ablation study results showing the performance of our grey-box attackers (DCA and MIAttacker) when the knowledge state vector is withheld (w/o \texttt{kstate\_emb}). This setup degrades them to black-box attackers. All experiments are conducted with a 10\% forgetting ratio. \textbf{Bold} values indicate the best performance for each model-dataset block under this ablated setting.}
\label{ab_01}
\end{table*}
\begin{table*}[ht]
\centering
\renewcommand{\arraystretch}{1.2}
\resizebox{\textwidth}{!}{%
\begin{tabular}{cc|cccc|cccc|cccc}
\hline
\multicolumn{2}{c|}{\textbf{Dataset}} &
  \multicolumn{4}{c|}{\textbf{Math1}} &
  \multicolumn{4}{c|}{\textbf{Math2}} &
  \multicolumn{4}{c}{\textbf{Frcsub}} \\ \hline
\multicolumn{2}{c|}{\textbf{Target w/ Defenses}} &
  \multicolumn{2}{c}{\textbf{\begin{tabular}[c]{@{}c@{}}DCA w/o\\ kstate\_emb\end{tabular}}} &
  \multicolumn{2}{c|}{\textbf{\begin{tabular}[c]{@{}c@{}}MIAttacker w/o\\ kstate\_emb\end{tabular}}} &
  \multicolumn{2}{c}{\textbf{\begin{tabular}[c]{@{}c@{}}DCA w/o\\ kstate\_emb\end{tabular}}} &
  \multicolumn{2}{c|}{\textbf{\begin{tabular}[c]{@{}c@{}}MIAttacker w/o\\ kstate\_emb\end{tabular}}} &
  \multicolumn{2}{c}{\textbf{\begin{tabular}[c]{@{}c@{}}DCA w/o\\ kstate\_emb\end{tabular}}} &
  \multicolumn{2}{c}{\textbf{\begin{tabular}[c]{@{}c@{}}MIAttacker w/o\\ kstate\_emb\end{tabular}}} \\ \hline
\textbf{Target Model} &
  \textbf{Defenses} &
  $\text{ACC}_{\text{MIA}}$ &
  $\text{AUC}_{\text{MIA}}$ &
  $\text{ACC}_{\text{MIA}}$ &
  $\text{AUC}_{\text{MIA}}$ &
  $\text{ACC}_{\text{MIA}}$ &
  $\text{AUC}_{\text{MIA}}$ &
  $\text{ACC}_{\text{MIA}}$ &
  $\text{AUC}_{\text{MIA}}$ &
  $\text{ACC}_{\text{MIA}}$ &
  $\text{AUC}_{\text{MIA}}$ &
  $\text{ACC}_{\text{MIA}}$ &
  $\text{AUC}_{\text{MIA}}$ \\ \hline
\multirow{4}{*}{NeuralCD} &
  * &
  \textbf{0.5236} &
  \textbf{0.5977} &
  0.5039 &
  0.5538 &
  \textbf{0.6024} &
  \textbf{0.5988} &
  \textbf{0.5542} &
  \textbf{0.5736} &
  \textbf{0.7381} &
  \textbf{0.8614} &
  0.6190 &
  \textbf{0.7727} \\
 &
  Amnesiac &
  \textbf{0.5236} &
  0.5968 &
  \textbf{0.5827} &
  \textbf{0.5791} &
  0.5743 &
  0.5959 &
  0.5502 &
  0.5294 &
  \textbf{0.7381} &
  0.8591 &
  0.5714 &
  0.7273 \\
 &
  L-CODEC &
  0.3228 &
  0.3300 &
  0.4252 &
  0.3922 &
  0.3976 &
  0.3381 &
  0.3614 &
  0.3512 &
  \textbf{0.7381} &
  \textbf{0.8614} &
  \textbf{0.6667} &
  0.7568 \\
 &
  SSD &
  0.5157 &
  0.5102 &
  0.4882 &
  0.5168 &
  0.4859 &
  0.4752 &
  0.4819 &
  0.4891 &
  0.6190 &
  0.6750 &
  0.5714 &
  0.6273 \\ \hline
\multirow{4}{*}{KSCD} &
  * &
  0.6535 &
  \textbf{0.6871} &
  \textbf{0.6732} &
  \textbf{0.6703} &
  \textbf{0.5502} &
  \textbf{0.5803} &
  \textbf{0.5582} &
  \textbf{0.5603} &
  \textbf{0.6905} &
  \textbf{0.7932} &
  0.5476 &
  0.6500 \\
 &
  Amnesiac &
  \textbf{0.6614} &
  0.6868 &
  0.5591 &
  0.6134 &
  0.5462 &
  0.5734 &
  0.5341 &
  0.5474 &
  \textbf{0.6905} &
  \textbf{0.7932} &
  0.5952 &
  0.6500 \\
 &
  L-CODEC &
  0.5039 &
  0.4856 &
  0.5354 &
  0.4938 &
  0.3293 &
  0.3142 &
  0.3976 &
  0.3463 &
  0.5238 &
  0.5159 &
  \textbf{0.7619} &
  \textbf{0.7273} \\
 &
  SSD &
  0.5472 &
  0.5387 &
  0.5748 &
  0.5483 &
  0.4900 &
  0.5243 &
  0.5462 &
  0.5355 &
  0.2857 &
  0.2727 &
  0.3810 &
  0.4568 \\ \hline
\multirow{4}{*}{KaNCD} &
  * &
  \textbf{0.5787} &
  0.6187 &
  \textbf{0.5787} &
  \textbf{0.6142} &
  \textbf{0.5141} &
  \textbf{0.5535} &
  \textbf{0.5422} &
  \textbf{0.5410} &
  \textbf{0.6905} &
  \textbf{0.8432} &
  \textbf{0.6190} &
  \textbf{0.7977} \\
 &
  Amnesiac &
  \textbf{0.5787} &
  \textbf{0.6216} &
  0.5512 &
  0.6104 &
  \textbf{0.5141} &
  0.5519 &
  0.5261 &
  0.5338 &
  \textbf{0.6905} &
  \textbf{0.8432} &
  \textbf{0.6190} &
  0.7273 \\
 &
  L-CODEC &
  0.3661 &
  0.3122 &
  0.3504 &
  0.3274 &
  0.3293 &
  0.2865 &
  0.4257 &
  0.4221 &
  0.3095 &
  0.3045 &
  0.2857 &
  0.4159 \\
 &
  SSD &
  0.5394 &
  0.5263 &
  0.5079 &
  0.5055 &
  0.4016 &
  0.4585 &
  0.5863 &
  0.5170 &
  0.3810 &
  0.4068 &
  0.4048 &
  0.4114 \\ \hline
\end{tabular}%
}
\caption{Ablation study results showing the performance of our grey-box attackers (DCA and MIAttacker) when the knowledge state vector is withheld (w/o \texttt{kstate\_emb}). This setup degrades them to black-box attackers. All experiments are conducted with a 1\% forgetting ratio. \textbf{Bold} values indicate the best performance for each model-dataset block under this ablated setting.}
\label{ab_001}
\end{table*}
\section{Supplementary Information on Experimental Results}
\subsection{Detailed Analysis of Main Results Across Unlearning Ratios}
The complete results presented in Table~\ref{mainres_01}, Table~\ref{mainres}, and Table~\ref{mainres_001} allow for a deeper analysis of the P-MIA framework's performance across different unlearning ratios (10\%, 5\%, and 1\%). This analysis further reinforces and adds nuance to the conclusions drawn in the main paper. \textbf{Note that the ACC and AUC metrics in Tables~\ref{mainres_01} and \ref{mainres_001}, consistent with Table~\ref{mainres}, refer specifically to the MIA performance ($\text{ACC}_{\text{MIA}}$ and $\text{AUC}_{\text{MIA}}$).
}
\begin{figure*}[ht]
		\centering
		\begin{minipage}[t]{0.48\textwidth}
			\centering
			\includegraphics[width=8.5cm]{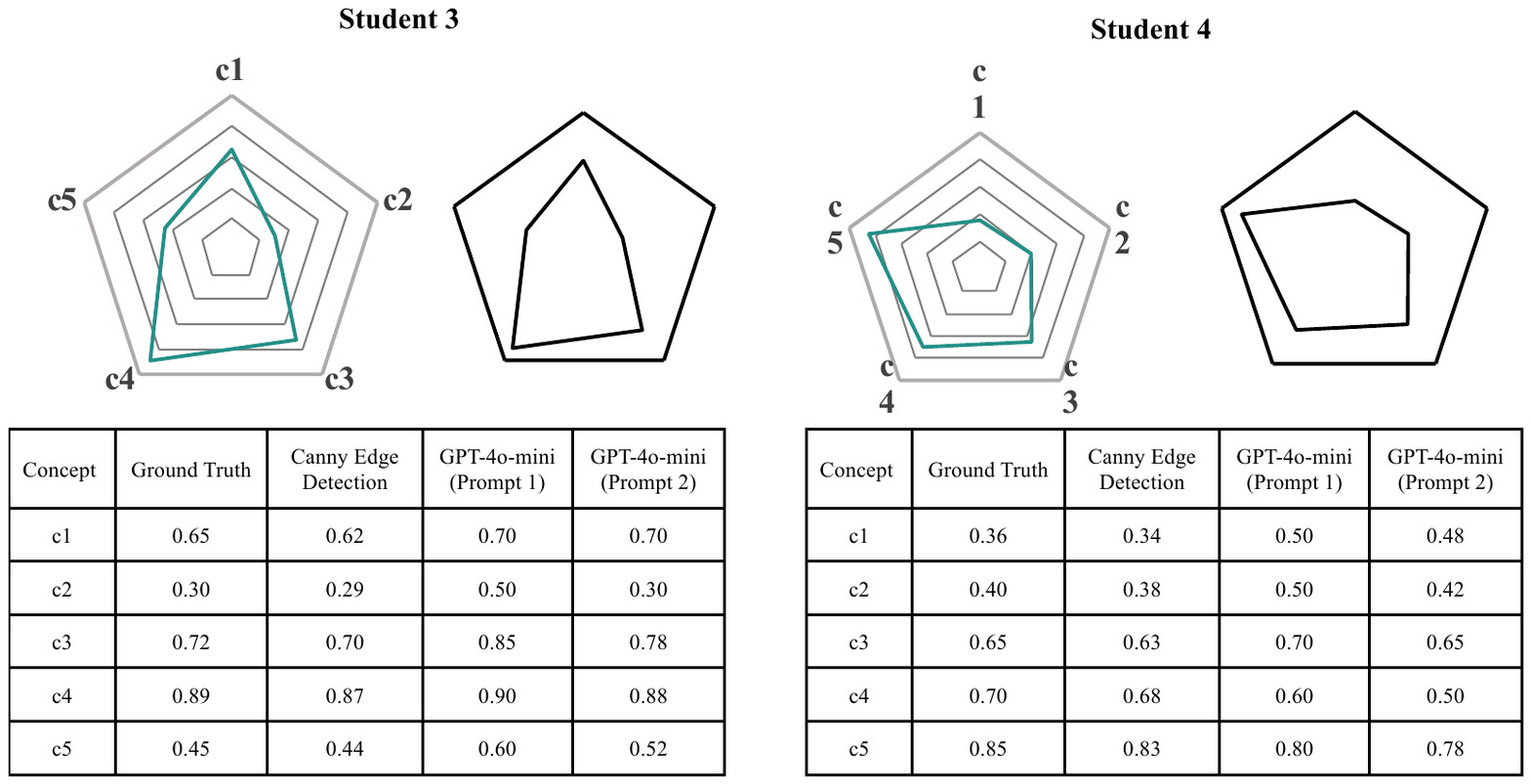}
			\subcaption{Example Students 3 and 4}
		\end{minipage}
		\begin{minipage}[t]{0.48\textwidth}
			\centering
			\includegraphics[width=8.5cm]{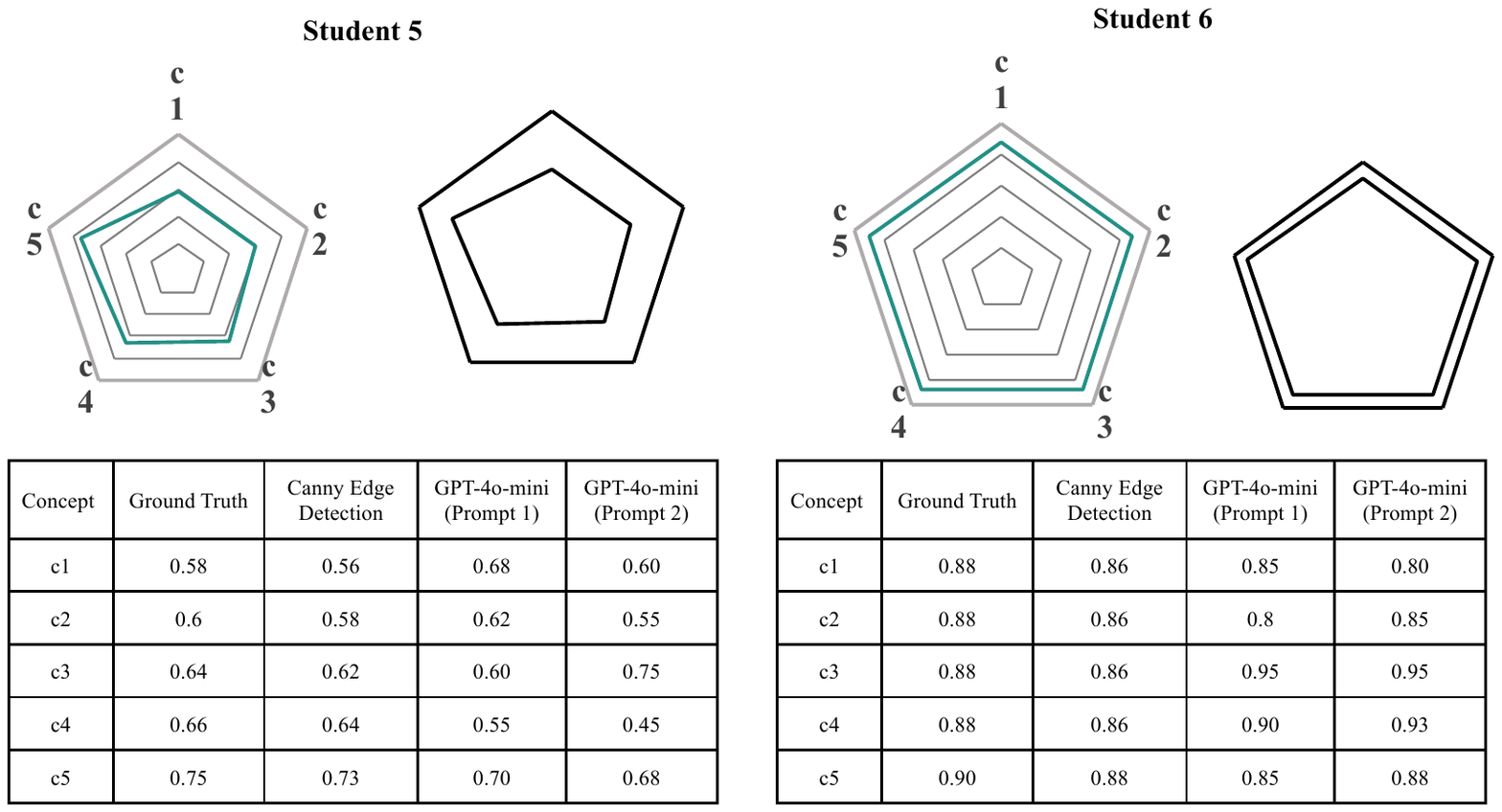}
			\subcaption{Example Students 5 and 6}
		\end{minipage}
		\caption{Additional validation results for the automated extraction of \texttt{kstate\_emb} from radar charts for four more representative students (Students 3-6). The results consistently show the high accuracy of the Canny Edge Detection method and the viability of using LLMs for the same purpose.}
		\label{vs2}
\end{figure*}
\subsubsection{Robustness of Core Findings}Across all tested unlearning ratios, our two primary findings from the main paper remain unequivocally robust:
\begin{itemize}
    \item \textbf{Superiority of Grey-Box Attacks:} The performance gap between our grey-box attackers (DCA and MIAttacker) and the black-box (XGBoost) baseline is consistently massive, regardless of the percentage of data being forgotten. This confirms that the privacy risk posed by the exposed \texttt{kstate\_emb} is a fundamental vulnerability.
    \item \textbf{Insufficiency of Approximate Unlearning Defenses:} While all approximate unlearning methods provide some degree of mitigation, they consistently fail to reduce the attack success rate to the near-random level of the ideal ''Retrain'' model. A significant privacy risk persists in all tested scenarios against grey-box attackers.
\end{itemize}
\begin{figure*}[ht]
		\centering
		\begin{minipage}[t]{0.31\textwidth}
			\centering
			\includegraphics[width=5.5cm]{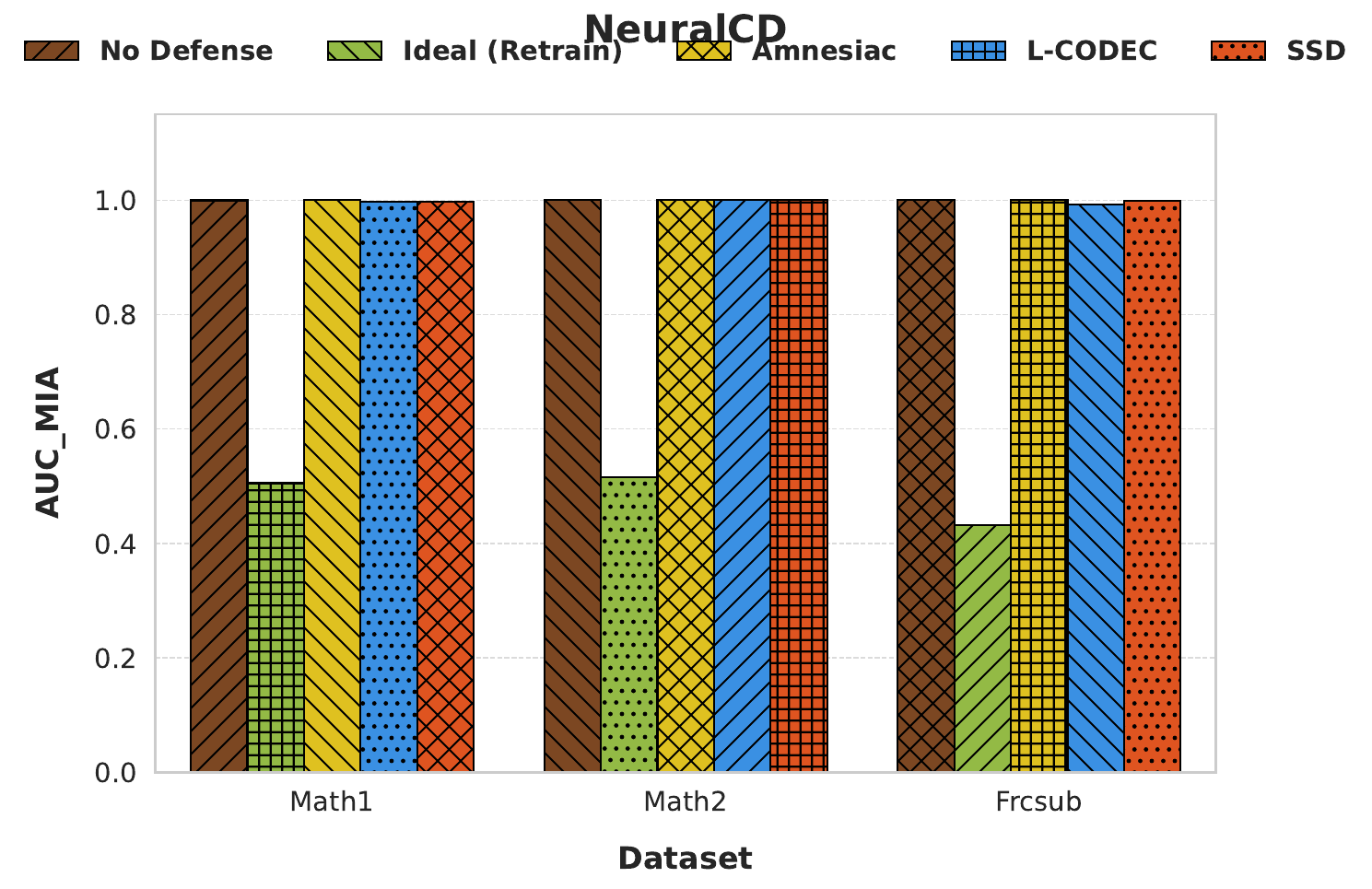}
			\subcaption{NeuralCD}
		\end{minipage}
		\begin{minipage}[t]{0.31\textwidth}
			\centering
			\includegraphics[width=5.5cm]{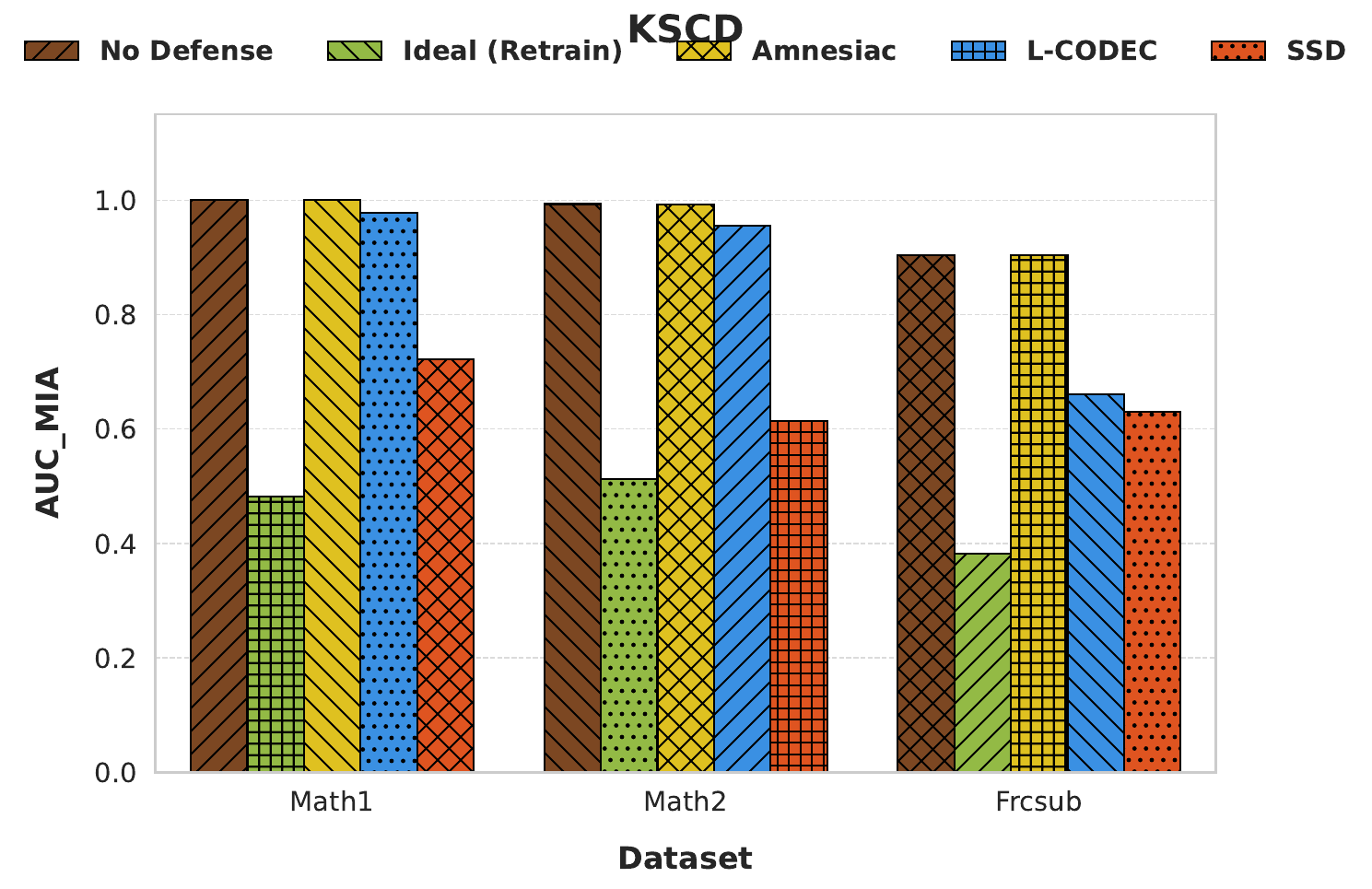}
			\subcaption{KSCD}
		\end{minipage}
		\begin{minipage}[t]{0.31\textwidth}
			\centering
			\includegraphics[width=5.5cm]{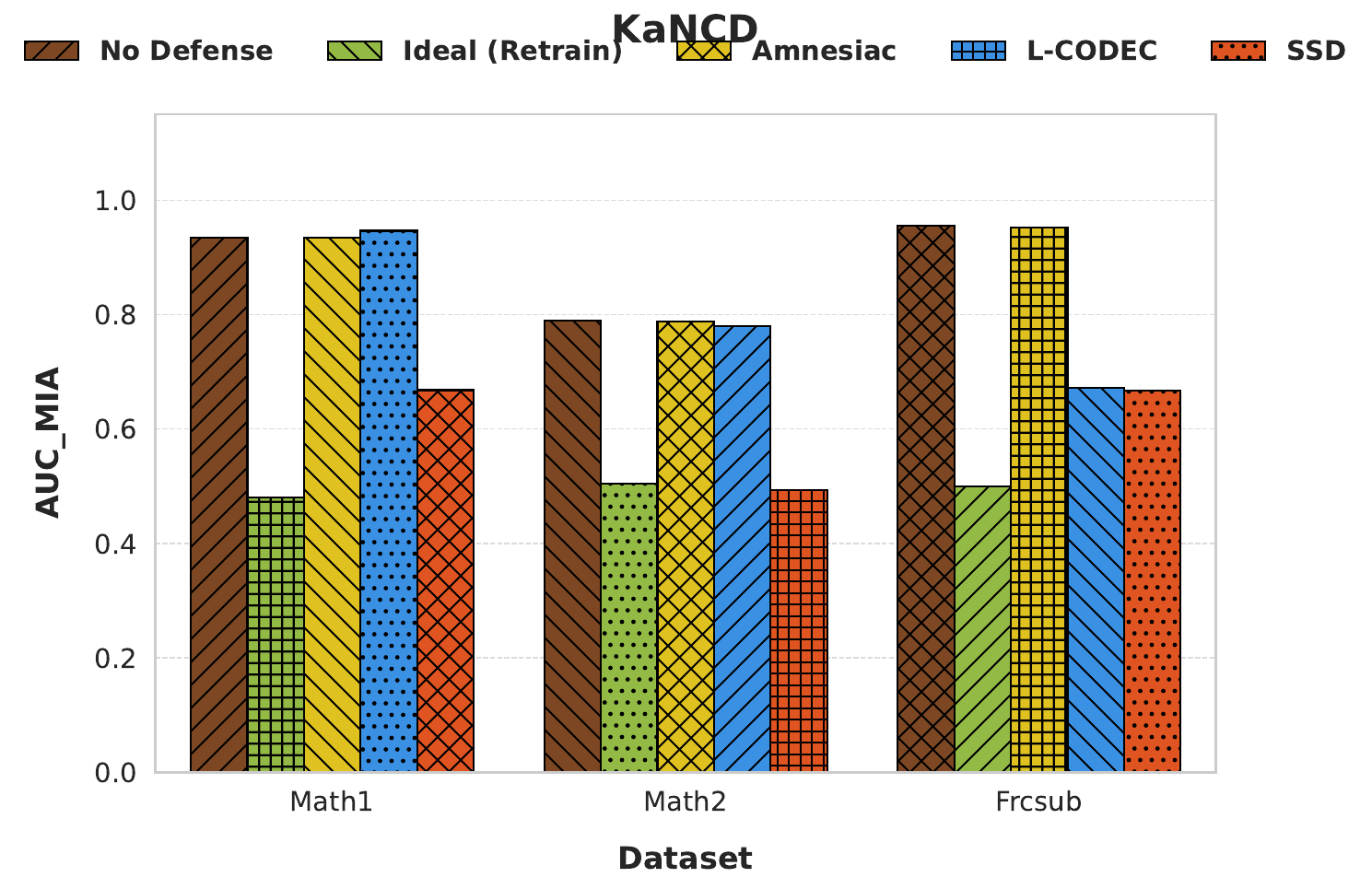}
			\subcaption{KaNCD}
		\end{minipage}
		\caption{Audit of machine unlearning defenses using our P-MIA (MIAttacker) with a 10\% forgetting ratio. Each subplot shows the attack AUC\textsubscript{MIA} against a different target CDM.}
		\label{vs2}
\end{figure*}
\begin{figure*}[ht]
		\centering
		\begin{minipage}[t]{0.31\textwidth}
			\centering
			\includegraphics[width=5.5cm]{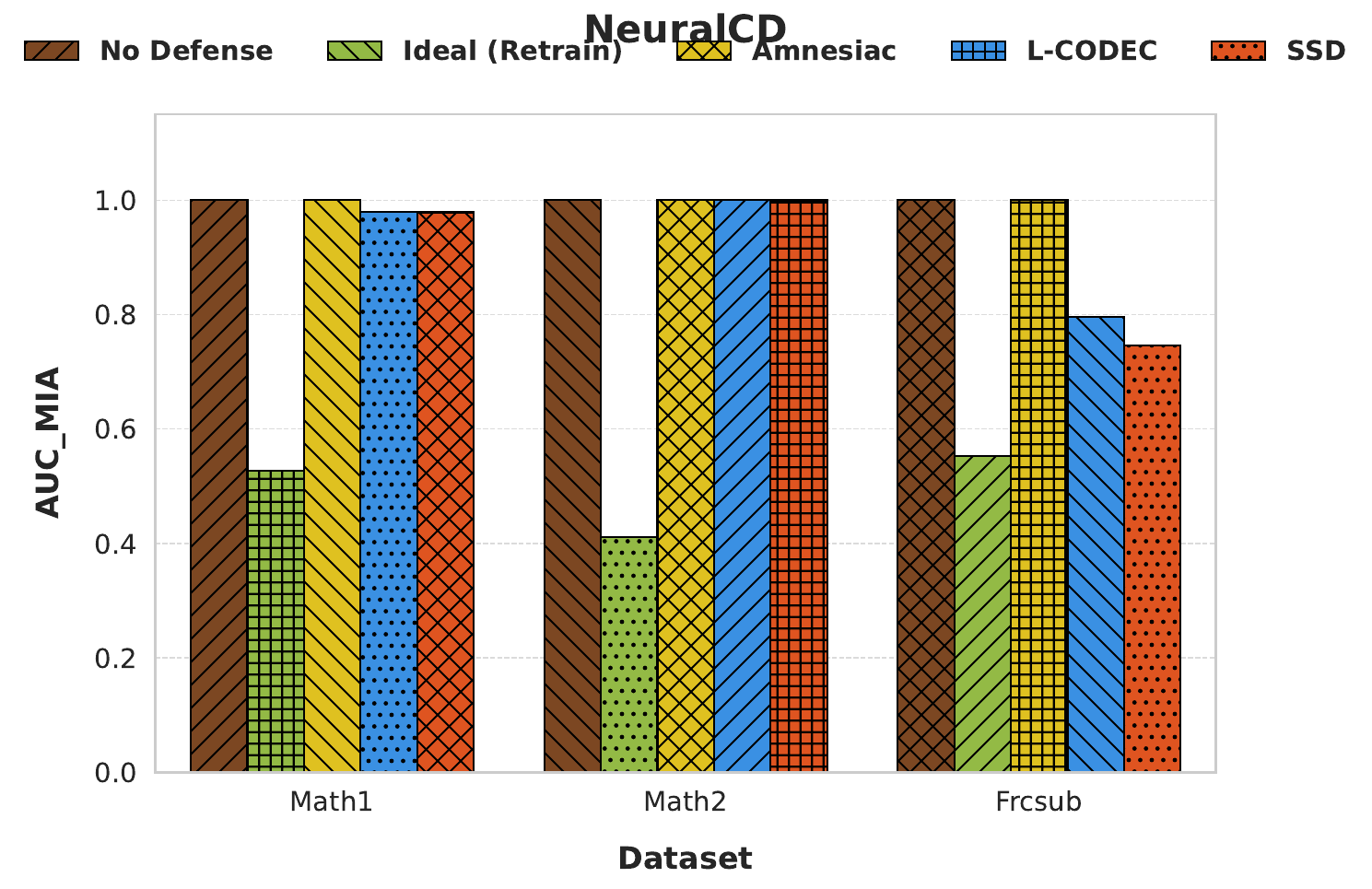}
			\subcaption{NeuralCD}
		\end{minipage}
		\begin{minipage}[t]{0.31\textwidth}
			\centering
			\includegraphics[width=5.5cm]{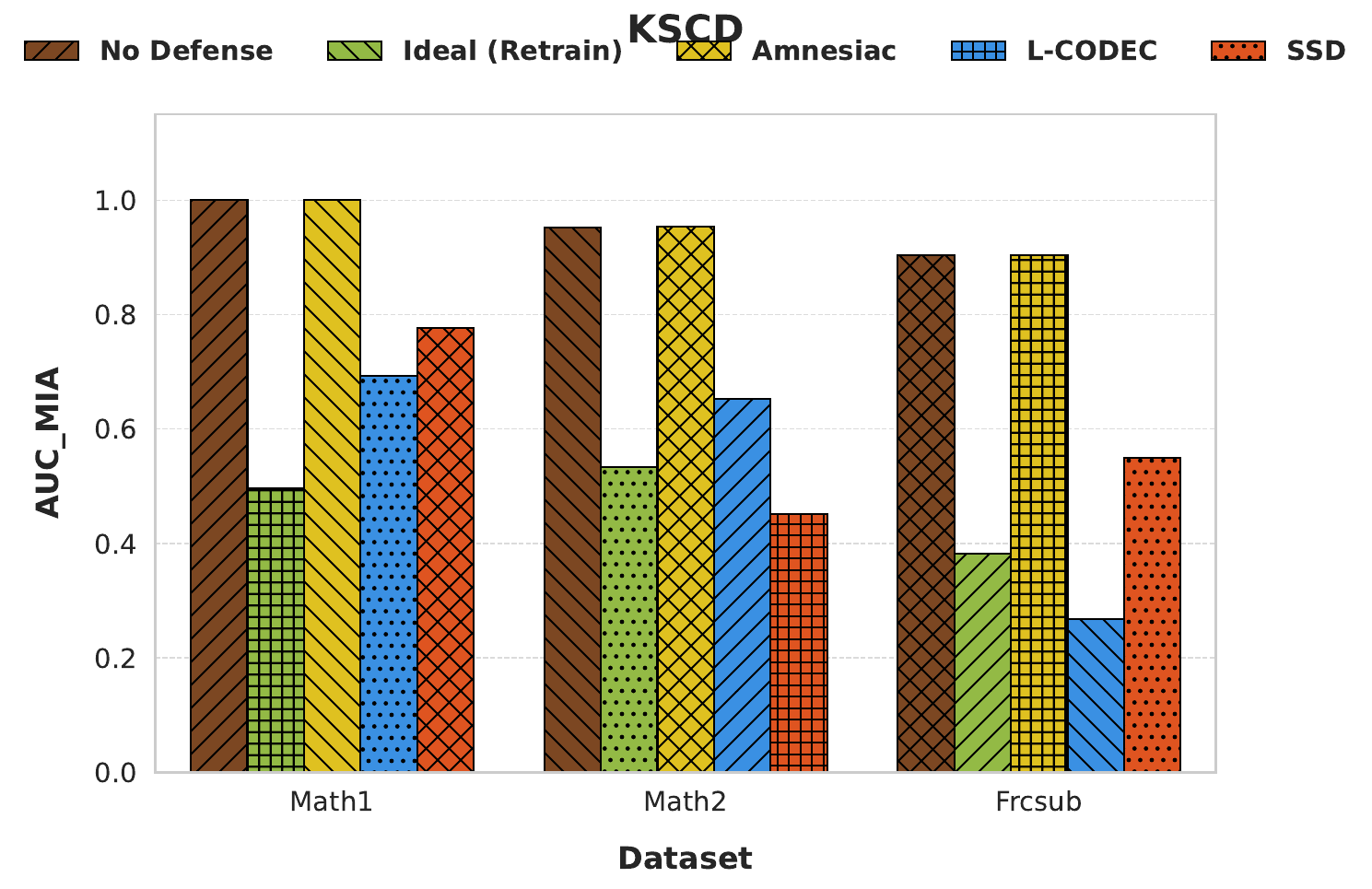}
			\subcaption{KSCD}
		\end{minipage}
		\begin{minipage}[t]{0.31\textwidth}
			\centering
			\includegraphics[width=5.5cm]{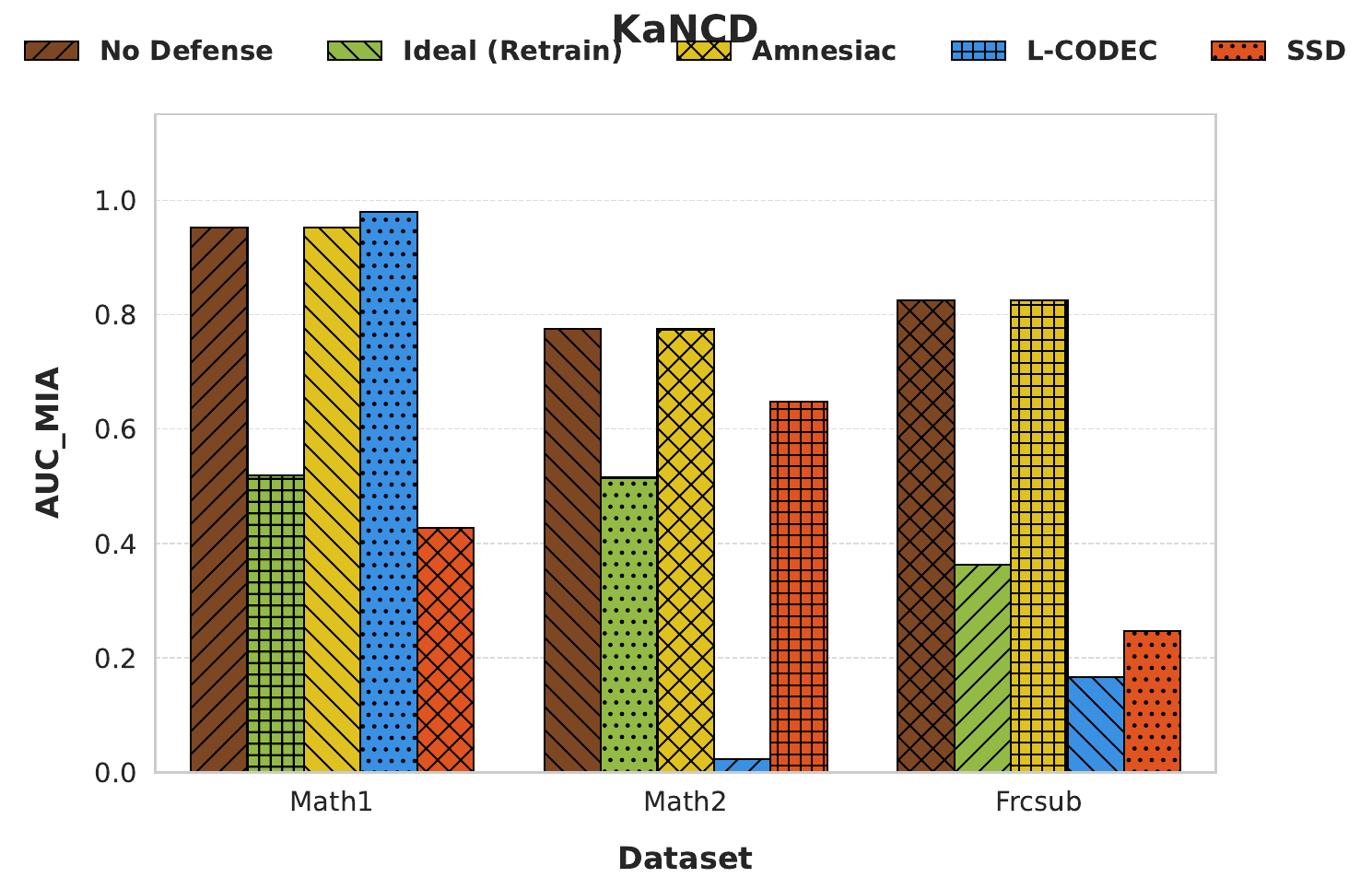}
			\subcaption{KaNCD}
		\end{minipage}
		\caption{Audit of machine unlearning defenses using our P-MIA (MIAttacker) with a 1\% forgetting ratio. Each subplot shows the attack AUC\textsubscript{MIA} against a different target CDM.}
		\label{vs3}
\end{figure*}
\subsubsection{Impact of Unlearning Ratio on Defense Efficacy}A clear trend emerges from comparing the tables: the effectiveness of the approximate unlearning defenses is generally inversely correlated with the unlearning ratio. That is, defenses perform better (i.e., yield lower attack AUCs) when the fraction of data to be forgotten is smaller (1\%), and their efficacy diminishes as the ratio increases (10\%).
\begin{itemize}
    \item \textbf{Quantitative Evidence:} For instance, when defending the KSCD model on the Frcsub dataset with the SSD method, the MIAttacker's AUC systematically increases with the forgetting ratio: it is \textbf{0.4587} at the 1\% ratio, remains at \textbf{0.4587} for the 5\% ratio, and rises to \textbf{0.5829} at the 10\% ratio. This pattern is broadly consistent across most experiments.
    \item \textbf{Rationale:} This outcome is logical. Approximate unlearning algorithms must perturb the model to erase information. Erasing the influence of a larger data slice (10\%) requires a more substantial and potentially less precise perturbation, leaving more residual fingerprints for our sensitive P-MIA framework to exploit. Conversely, forgetting a smaller, more targeted set of data (1\%) can be done more ``surgically'', resulting in a model state that is closer to the ideal retrained model.
\end{itemize}

\subsubsection{Impact of Unlearning Ratio on Baseline Vulnerability}For the undefended models (`*`), we observe that the baseline vulnerability remains at a consistently high level across all unlearning ratios. For example, the MIAttacker AUC against the undefended KaNCD model on Math1 was 0.9328 (10\% ratio), 0.9398 (5\% ratio), and 0.9331 (1\% ratio). The minor fluctuations are not significant, indicating that the model's inherent vulnerability is a fundamental characteristic, largely independent of the size of the member set chosen for evaluation.

In summary, while the unlearning ratio modulates the precise degree of defense efficacy, it does not alter our core conclusions. The experiments across all ratios consistently demonstrate that (a) learner profiles are a critical privacy vulnerability, and (b) our P-MIA framework is a robust tool for demonstrating that current general-purpose unlearning defenses are not a panacea, especially as the scale of data removal increases.

\subsection{Quantifying the Impact of the Knowledge State Vector}
To empirically validate our central hypothesis, that the exposed knowledge state vector (\texttt{kstate\_emb}) is the primary driver of the grey-box attack's power—we conducted a critical ablation study. In this study, we evaluated the performance of our grey-box attackers (DCA and MIAttacker) when the \texttt{kstate\_emb} feature was withheld. Tables \ref{abres}, \ref{ab_01} and \ref{ab_001} present the results for this ``knowledge-blind'' configuration across different unlearning ratios. A direct comparison with the full grey-box attack results in Appendix~B.1 provides the following conclusive insights.

\subsubsection{The \texttt{kstate\_emb} as the Core Source of Attack Performance}The most significant finding is the dramatic and consistent performance collapse of both DCA and MIAttacker across all datasets, target models, defense scenarios, and unlearning ratios once the \texttt{kstate\_emb} is removed. This performance degradation is systematic, proving that \texttt{kstate\_emb} is not merely a supplementary feature.
\begin{itemize}
    \item \textbf{Quantitative Comparison (5\% Ratio):} When attacking the undefended KSCD model on the Math1 dataset, the MIAttacker's AUC plummets from \textbf{1.0000} (with \texttt{kstate\_emb}, see Table \ref{mainres} down to a mere \textbf{0.6335} in the ablated setting (Table \ref{abres}). Similarly, against the L-CODEC defended KaNCD model on Math2, the MIAttacker's AUC drops from \textbf{0.9879} to just \textbf{0.3056}.
    \item \textbf{Consistency Across Ratios:} This massive performance gap persists at the 10\% and 1\% ratios. For instance, at the 10\% ratio, attacking the undefended NeuralCD on Math1 sees the DCA's AUC fall from \textbf{0.9996} to \textbf{0.7575} (Table \ref{ab_01}). At the 1\% ratio, attacking the undefended KSCD on Frcsub, the MIAttacker's AUC drops from \textbf{0.9595} to \textbf{0.6500} (Table \ref{ab_001}).
\end{itemize}

\subsubsection{Attack Degradation to Black-Box Levels}Without the \texttt{kstate\_emb}, the DCA and MIAttacker are effectively reduced to attackers using only $[\texttt{predict\_proba}, \texttt{response}]$ features, and their performance should thus be comparable to our black-box baseline (XGBoost). The results confirm this.
\begin{itemize}
    \item \textbf{Performance Comparison:} In the 5\% ablation study (Table \ref{abres}), the DCA's AUC on the undefended NeuralCD (Math1) is \textbf{0.7073}. This is in the same performance bracket as the XGBoost baseline's AUC of \textbf{0.8194} from the main experiment (Table \ref{mainres}) and significantly lower than the full grey-box attack's \textbf{1.0000}. This shows that without the \texttt{kstate\_emb}, even a sophisticated neural network attacker cannot surpass the information ceiling of the black-box baseline.
\end{itemize}

In summary, this comprehensive ablation study provides unequivocal evidence for our core thesis. The \texttt{kstate\_emb} is not just key to improving the attack; it is the \textbf{dominant source} of the exploited privacy leakage. Its removal cripples the grey-box attack, confirming that the learner profile itself constitutes a severe, standalone vulnerability. Access to this internal state information is the fundamental reason that elevates our P-MIA from a standard attack to a significantly more potent threat.
\subsection{Auditing Machine Unlearning Defenses with P-MIA}
Figure~\ref{vs} in the main paper presents the extraction results for two representative student examples. To demonstrate the consistency and robustness of our methods, we provide additional results for four more randomly selected students in Figure~\ref{vs2}. The MAE for our Canny Edge Detection method consistently remains in the low range of 0.01--0.03, confirming its high fidelity. The LLM-based approaches also provide reasonable estimates, validating that multiple pathways exist for an attacker to reliably obtain the \texttt{kstate\_emb} from public-facing visualizations.

The prompt1 and prompt2 used in Tables \ref{vs} and \ref{vs2} are shown as below.
\begin{tcolorbox}[
    colback=white,      
    colframe=black,     
    boxrule=0.5pt,      
    arc=2mm,            
    title=\textbf{Prompt 1: General Instruction}, 
    fonttitle=\bfseries
]
In cognitive diagnosis (CD) tasks, student knowledge states are often visualized using radar charts, where the student’s mastery level for each concept is represented as a value between 0 and 1. The green line in the radar chart indicates the estimated mastery levels for five specific concepts, while the gray lines serve as reference levels, corresponding to the values 0.2, 0.4, 0.6, 0.8, and 1.0.
\newline\newline
Based on the relative position of the green line with respect to the reference lines, estimate the student’s mastery level for each concept on a continuous scale from 0 to 1.
\end{tcolorbox}

\begin{tcolorbox}[
    colback=white,
    colframe=black,
    boxrule=0.5pt,
    arc=2mm,
    title=\textbf{Prompt 2: In-Context Learning Example},
    fonttitle=\bfseries
]
In cognitive diagnosis (CD) tasks, radar charts are commonly used to visualize a student’s mastery level across several concepts. Each concept is represented as a vertex, and the student’s mastery is quantified as a value between 0 and 1. In the radar chart, the green line indicates the student’s estimated mastery levels, while the five gray concentric pentagons represent reference levels corresponding to 0.2, 0.4, 0.6, 0.8, and 1.0, respectively. For example, in the first radar chart: The vertex for c1 lies between the third (0.6) and fourth (0.8) reference lines, approximately midway between them, suggesting a value of 0.65. The vertex for c2 lies exactly on the second reference line, suggesting a value of 0.40. The vertex for c3 lies between the third and fourth lines, close to the fourth, indicating a value of 0.79. The vertex for c4 lies between the fourth and fifth lines, closer to the fifth, implying a value of 0.92. The vertex for c5 lies between the fourth and fifth lines, but closer to the fourth, indicating a value of 0.83.
Based on this interpretation approach, estimate the mastery levels for the student in the second radar chart by visually comparing the positions of the green line with respect to the gray reference lines. Report each concept’s estimated value on a continuous scale from 0 to 1.
\end{tcolorbox}

While the main paper primarily reports the MIA audit results for the 5\% forgetting ratio, this section provides the complete figures for the 10\% (Figure~\ref{vs2}) and 1\% (Figure~\ref{vs3}) scenarios, followed by an in-depth cross-ratio analysis.

First, the core conclusions drawn in the main paper—namely, the sensitivity of P-MIA as an audit tool and the insufficiency of approximate unlearning defenses—are fully validated across both the 10\% and 1\% scenarios. Regardless of the ratio, the AUC for undefended models remains near 1.0, the AUC for ideally retrained models is near 0.5, and all approximate methods fail to reduce the attack AUC to this ideal level. This confirms the generalizability of our findings.

\subsubsection{Inverse Correlation Between Unlearning Ratio and Defense Efficacy}By comparing the figures for all three ratios, a clear trend emerges: the effectiveness of approximate unlearning defenses is strongly and inversely correlated with the forgetting ratio. Defenses are significantly more effective when less data needs to be forgotten.
\begin{itemize}
    \item \textbf{Quantitative Evidence:} A clear example is the L-CODEC defense on the KaNCD model with the Frcsub dataset. At a \textbf{10\%} forgetting ratio, the attack AUC is \textbf{0.66} (Figure~\ref{vs2}(c)). This drops to \textbf{0.469} at the \textbf{5\%} ratio (Figure~\ref{vs}(c)), and further down to \textbf{0.3705} at the \textbf{1\%} ratio (Figure~\ref{vs3}(c)), where the attack is completely nullified and performs worse than random guessing.
    \item \textbf{General Trend:} This pattern—higher ratio leading to worse defense (higher attack AUC)—holds true for the vast majority of experimental configurations. For instance, defending KSCD on Math2 with SSD, the attack AUC increases from \textbf{0.5359} (1\%) to \textbf{0.6085} (5\%) and finally to \textbf{0.6128} (10\%).
\end{itemize}

\subsubsection{Observations at the Extremes}
\begin{itemize}
    \item \textbf{Challenges at 10\% Ratio:} At the high 10\% ratio, some defenses become almost entirely ineffective. For example, the Amnesiac defense against NeuralCD yields an attack AUC of nearly 1.0 across all three datasets, almost identical to the ``No Defense'' case. This suggests that simple gradient ascent struggles to handle larger-scale data removal.
    \item \textbf{Opportunities at 1\% Ratio:} At the low 1\% ratio, we observe instances where a defense can outperform ideal retraining (AUC < 0.5), such as the L-CODEC on Frcsub mentioned above. This may indicate an ``over-correction'' phenomenon, where the algorithm's perturbations not only remove member information but also introduce noise that leads the attack model to make systematic, inverted errors.
\end{itemize}

In summary, this comprehensive analysis across different unlearning ratios not only validates the robustness of our main conclusions but also reveals that the forgetting ratio is a key factor influencing the success of approximate unlearning. Our P-MIA framework accurately captures this variance in defense efficacy, further demonstrating its power as a fine-grained privacy auditing tool.
\end{document}